\documentclass[12pt,a4paper,english,twoside]{article}

\usepackage{tikz}
\usetikzlibrary{matrix,chains,positioning,decorations.pathreplacing,arrows}

\usepackage{multirow}
\usepackage{amsmath, amssymb, setspace} %Cosas de mates y fijar espaciado entre líneas
\usepackage{babel} %Elige el idioma de las palabras que se generan automáticamente (como tabla o Figura)
\usepackage[utf8]{inputenc} %Permite introducir directamente acentos: á en lugar de \'a etc.
\usepackage{graphicx} %Cosas chachis para Figuras
\usepackage{caption} %Cosas chachis en los nombres de las figuras
\usepackage{subfig}
\usepackage{float} %Permite manejar/mover/partir mejor figuras y tablas
\usepackage{fancyhdr} %Encabezado y pie de página chachi
\usepackage{mathtools} %Más símbolos matemáticos
\usepackage[toc,page]{appendix} %Incluir apéndices
\usepackage{rotating} %Para girar floating objects (texto en mbox, tablas, figuras...)
\usepackage{multicol} %Para dividir la página en columnas
\usepackage{verbatim} % comentarios
\usepackage{braket} %Permite usar bra-kets para la notación de Dirac
\usepackage{pdfpages} %Para añadir páginas de pdf externos
\usepackage{cancel} %Para tachar elementos en las fórmulas
\usepackage[nottoc]{tocbibind} %Pone LOF, LOT, Apendices en la TOC
\usepackage{titlesec} %Edit title format
\usepackage{slashed} %Dirac slash notation:  \slashed{p}
\usepackage{pgffor}							% For repeating patterns
\usepackage{booktabs}

\usepackage[pdftex, colorlinks=true,linkcolor=blue, urlcolor=blue, linktocpage]{hyperref}
\titleformat
{\chapter} % command
[display] % shape
{\bfseries\Large} % format
{  } % label
{-1cm} % sep
{ \thechapter.  } % before-code
[ \vspace{0.5ex} ] % after-code

\begin{document}
\renewcommand{\tablename}{\bf Table} %Cambia la palabra "Cuadro" por "Tabla"
\renewcommand{\figurename}{\bf Figure}
\renewcommand{\listtablename}{List of Tables} %Cambia el título "Índice de cuadros" por "Índice de Tablas
\renewcommand*\contentsname{Table of Contents} %Cambia el título "Índice General" por "Índice"
\renewcommand \listfigurename{List of Figures} %Cambia el título "Índice de figuras" por "Índice de Figuras"
\renewcommand{\appendixname}{Appendixes}
\renewcommand{\appendixtocname}{Appendixes}
\renewcommand{\appendixpagename}{Appendixes}
%\numberwithin{figure}{section}	%Estos comandos generan numeración Sección.Numero de Figura/tabla/ecuación
%\numberwithin{table}{section}
\numberwithin{equation}{section}
\renewcommand{\bibname}{References}

\def\nicefrac#1#2{\leavevmode%
    \raise.5ex\hbox{\small #1}%
    \kern-.1em/\kern-.15em%
    \lower.25ex\hbox{\small #2}}

%Estilo especial para pág. de principio de sección
\fancypagestyle{newstyle}{
\fancyhf{} % clear all header and footer fields
\fancyfoot[OR, EL]{\thepage} % except the center
\renewcommand{\headrulewidth}{0pt}
\renewcommand{\footrulewidth}{0pt}}

% Encabezado y pie de página
\pagestyle{fancy}
\renewcommand{\sectionmark}[1]{\markboth{#1}{}}
\renewcommand{\headrulewidth}{0pt}  %Grosor de la línea bajo el encabezado
\fancyhf{} %limpiar los campos
% página ... R=derecha, L=izquierda, C=centro; O=impar, E=par
% El número de la página es \thepage
%\fancyhead[RO]{\sectionmark}
%\fancyheadoffset{1cm}
\fancyfoot[LE]{\thepage}
\fancyfoot[RO]{\thepage}
%\fancyhead[CE]{\nouppercase{\thechaptr. \leftmark}}
\fancyhead[LO]{Section \thesection. \leftmark}
\fancyhead[RE]{J.M. Benlloch-Rodríguez}

\includepdf[pages={1}]{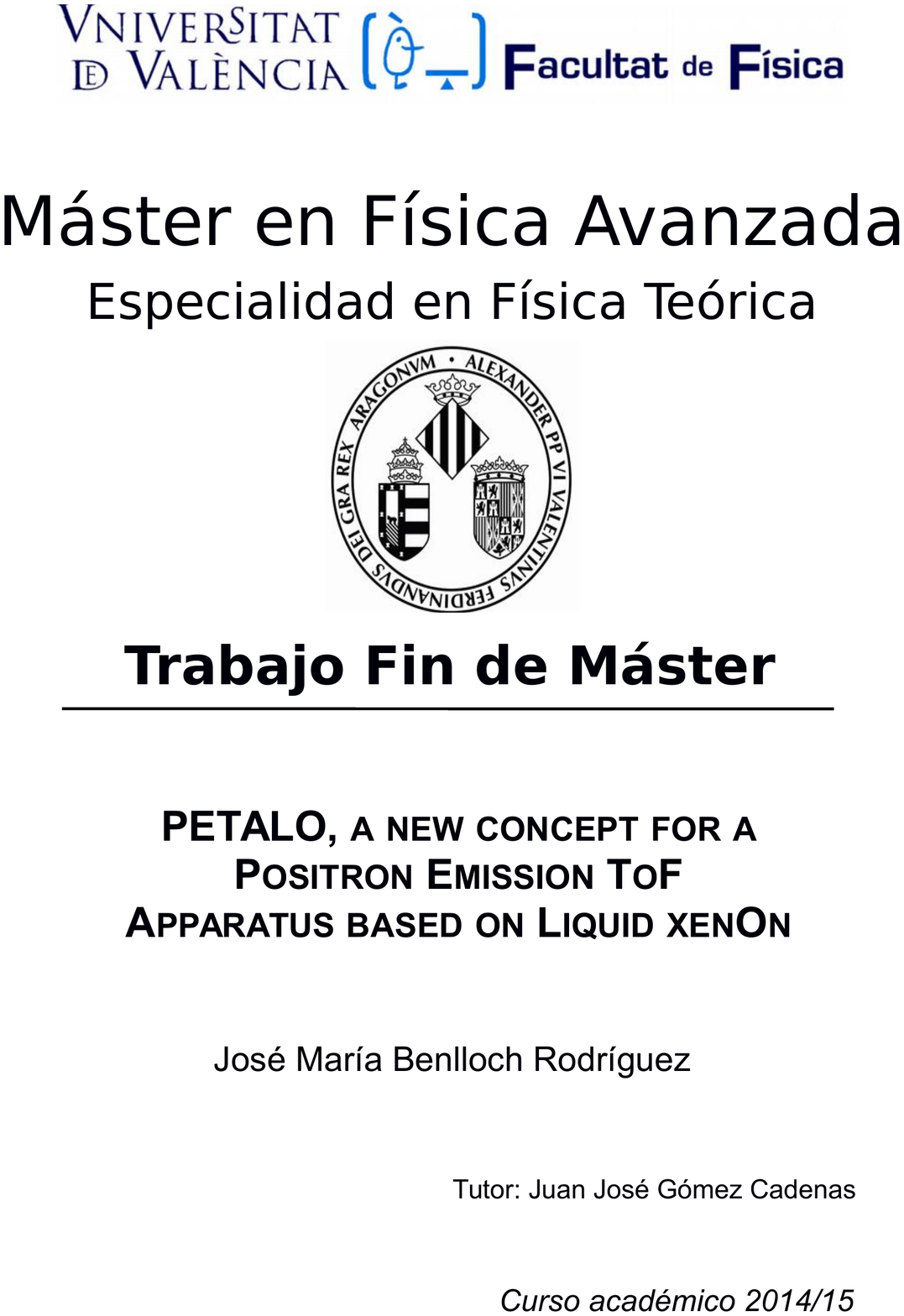}

\newpage\null\thispagestyle{empty}\newpage

\newpage
\thispagestyle{newstyle}

\section*{Abstract}
This master thesis presents a new type of Positron Emission TOF Apparatus using Liquid xenOn (PETALO). The detector is based in the  Liquid Xenon Scintillating Cell (LXSC).  The cell is a box filled with liquid xenon (LXe) whose transverse dimensions are chosen to optimize packing and with a thickness optimized to contain a large fraction of the incoming photons. The entry and exit faces of the box (relative to the incoming gammas direction) are instrumented with large silicon photomultipliers (SiPMs), coated with a wavelength shifter, tetraphenyl butadiene (TPB). The non-instrumented faces are covered by reflecting Teflon coated with TPB. In this thesis we show that the LXSC can display an energy resolution of 5\% FWHM, much better than that of conventional solid scintillators such as LSO/LYSO. The LXSC can measure the interaction point of the incoming photon with a resolution in the three coordinates of 1 mm. The very fast scintillation time of LXe (2 ns) and the availability of suitable sensors and electronics permits a coincidence resolution time (CRT) in the range of 100-200 ps, again much better than any current PET-TOF system. The LXSC constitutes the core of a high-sensitivity, nuclear magnetic resonance compatible, PET device, with enhanced Time Of Flight (TOF) sensitivity.

\newpage\null\thispagestyle{empty}\newpage

\newpage
\thispagestyle{newstyle}
\tableofcontents

\newpage
\thispagestyle{newstyle}

\section{Introduction}
\label{sec.intro}

Positron Emission Tomography (PET) is a non invasive imaging technique that produces a three-dimensional image of functional processes in the body. The system detects pairs of gamma rays emitted indirectly by a positron-emitting radionuclide (tracer), which is introduced into the body on a biologically active molecule. 

PET technology has evolved rapidly during the last decade, thanks to the introduction of high-yield, high-resolution, and fast-response solid scintillator detectors, such as LSO/LYSO. At the same time, a new type of sensors, the so-called silicon-photomultipliers or SiPMs, are quickly replacing conventional PMTs as readout devices. SiPMs have large gains, comparable to that of PMTs, excellent photon detection efficiency (PDE), close to 50\% around 420 nm can be fabricated in a variety of small size dices, allowing the construction of very modular pixelated systems. Furthermore, the use of SiPMs makes PET compatible with technologies that require very intense magnetic fields, such as Magnetic Resonance Imaging (MRI). 

However, a limitation of this excellent technology is the high cost of the scanner. The driving factor is the high cost of crystals such as LSO or LYSO, but the large number of channels needed for good energy and spatial resolution are also relevant. 

This master thesis presents a new type of Positron Emission TOF Apparatus using Liquid xenOn (PETALO). The use of liquid xenon (LXe) improves the energy resolution (ER) and the coincidence resolution time (CRT) which can be achieved by conventional PET systems based in LSO/LYSO. At the same time, LXe is much cheaper than LSO/LYSO and the high yield and homogeneity of the liquid allows a sparser instrumentation. As a consequence, PETALO may show better performance at lower cost than conventional PET scanners, thus offering a potential break-through of the technology.  

PETALO is based in the  Liquid Xenon Scintillating Cell (LXSC).  The cell is a box filled with liquid xenon (LXe) whose transverse dimensions are chosen to optimize packing and with a thickness optimized to contain a large fraction of the incoming photons. The entry and exit faces of the box (relative to the incoming gammas direction) are instrumented with large silicon photomultipliers (SiPMs), coated with a wavelength shifter, tetraphenyl butadiene (TPB). The non-instrumented faces are covered by reflecting Teflon coated with TPB. 

In this thesis we show that the LXSC can display an energy resolution better than 5\% FWHM, much better than that of conventional solid scintillators such as LSO/LYSO. The LXSC can measure the interaction point of the incoming photon with a resolution in the three coordinates of 1--2 mm. The very fast scintillation time of LXe  and the availability of suitable sensors and electronics permits a CRT in the range of 100-200 ps, again much better than any current PET-TOF system. 

This document is organized as follows. In section \ref{sec.pet} we introduce the basic ideas of Positron Emission Tomography. In section \ref{sec.ppet}, we discuss the performance characteristics of PET scanners. Section \ref{sec.ssd} compares the conventional solid scintillating detectors used in conventional PET scanners with LXe. The PETALO concept is introduced in section \ref{sec.petalo}. Section \ref{sec.sipm} describes the concept and main performance parameters of silicon photomultipliers. The LXSC is discussed in section \ref{sec.lxsc} and the study of the LXSC properties is discussed in section \ref{sec.mc}. 
%In section \ref{sec.comp} we briefly comment on how to resolve photoelectric and Compton interactions. 
We sketch possible PETALO scanners in section \ref{sec.pets}. Finally in section \ref{sec.conclu} we present our conclusions. 

\section{Positron Emission Tomography}
\label{sec.pet}

\begin{figure}[!b]
	\centering
	\includegraphics[scale=0.8]{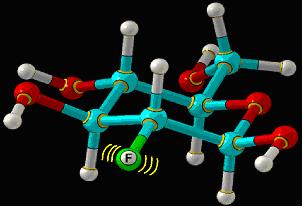}
	\includegraphics[scale=0.8]{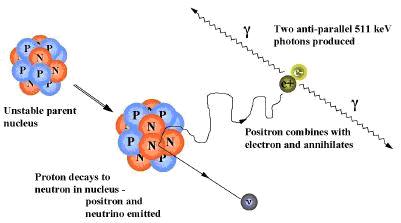}
	\caption{\label{fig.fdg}: Left panel, FDG is an organic molecule which contains a radioactive isotope of Fluor. The radionuclide decays emitting positrons, which annihilate after a short path length in the body tissue with an electron, resulting in the emission of two back-to-back photons of 511 keV (right panel). }
\end{figure}

A Positron Emission Tomography is a functional scan---it does not show anatomic features, but rather it measures metabolic activity of the cells of body tissues---. Used mostly in patients with brain or heart conditions and cancer, its big advantage is to identify the onset of a disease process before anatomical changes (that can be seen with other imaging processes such as computed tomography (CT) or MRI) related to the disease take place.

The PET technology  is based in the use of positron emitters radio-pharmaceuticals. 
If the biologically active molecule chosen is fluorodeoxyglucose (FDG), an analogue of glucose which contains a radioactive isotope of Fluor (F-18), the concentrations of tracer imaged will indicate tissue metabolic activity as it corresponds to the regional glucose uptake. Use of this tracer to explore the possibility of cancer metastasis (i.e., spreading to other sites) is the most common type of PET scan. However, many other radioactive tracers are used in PET to image the tissue concentration of other types of molecules of interest. 

Radioactive isotopes are produced (normally in a dedicated cyclotron) and attached to organic molecules that the studied cells absorb in their metabolism. The molecule and the radionuclide form the so called radiotracer. The tracer is injected to the patient where it will decay emitting a positron. The emitted positron travels a short distance before annihilating with an electron resulting in the emission of two 511 keV gamma rays in opposite directions (Figure \ref{fig.fdg}).

\begin{figure}[!bthp]
	\centering
	\includegraphics[scale=1.0]{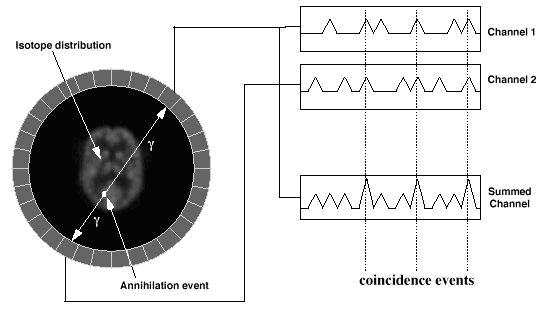}
	\caption{\label{fig.pet} Coincidence detection principle in a PET detector.}
\end{figure}

By detecting the two photons in coincidence and the coordinates of their interaction points in a rings of detectors surrounding the area under study is possible to define a line of response (LOR) along which the positron emitting source is located in the patient. A set of such intersecting lines allows 3D reconstruction of the source. The principle is illustrated in Figure \ref{fig.pet}.

\subsection{Medical applications of PET}

%\begin{figure}[!bthp]
%	\centering
%	\includegraphics[scale=2.0]{img/800px-PET-image.jpg}
%	\caption{\label{fig.brain} The image shows a transaxial slice of the brain of a 56 year old patient (male) taken with positron emission tomography (PET). The injected dose have been 282 MBq of FDG and the image was generated from a 20 minutes measurement with a ECAT Exact HR+ PET Scanner. Red areas show more accumulated tracer substance (FDG) and blue areas are regions where low to no tracer have been accumulated.}
%\end{figure}

The main applications of PET to medicine are oncology, neuroimaging and cardiology. The oncological applications are mostly based in the use of FDG, a glucose analog where an atom of fluorine-18 (F-18) replaces an atom of oxygen. A typical dose of FDG used in an oncological scan has an effective radiation dose of 14 mSv, equivalent to about 5 years of dose by background radiation (combining natural and artificial sources). It follows that one of the important issues in PET applications is to reduce the FDG dose as much as possible, which in turn requires to improve the performance of the PET scanners.

FDG is taken up by glucose-using cells and phosphorylated\footnote{Phosphorylation is the addition of a phosphate (PO$_4^{3--}$) group to a protein or other organic molecule. Phosphorylation and its counterpart, dephosphorylation, turn many protein enzymes on and off, thereby altering their function and activity.} by hexokinase\footnote{An enzyme that phosphorylates hexoses (six-carbon sugars), forming hexose phosphate. In most organisms, glucose is the most important substrate of hexokinases, and glucose-6-phosphate the most important product.}. Because the oxygen atom that is replaced by F-18 to generate FDG is required for the next step in glucose metabolism in all cells, no further reactions occur in FDG. Furthermore, most tissues cannot remove the phosphate added by hexokinase. This means that FDG is trapped in any cell that takes it up, until it decays. This results in intense radiolabeling of tissues with high glucose uptake, such as the brain, the liver, and most cancers. As a result, FDG-PET can be used for diagnosis, staging, and monitoring treatment of many types of cancer. The technology is also very well suited to search for tumor metastasis, or for recurrence after a known highly active primary tumor is removed. 

PET neuroimaging uses the fact that the brain is an avid user of glucose, and since brain pathologies such as Alzheimer's disease greatly decrease brain metabolism of glucose standard FDG-PET of the brain, which measures regional glucose use, may also be successfully used to differentiate Alzheimer's disease from other dementing processes, and also to make early diagnosis of Alzheimer's disease. 

In clinical cardiology, FDG-PET can identify so-called ``hibernating myocardium'', that is a state when some segments of the myocardium exhibit abnormalities of contractile function. These abnormalities can also be visualized with echocardiography, cardiac magnetic resonance imaging or ventriculography. FDG-PET imaging of atherosclerosis to detect patients at risk of stroke is also feasible and can help test the efficacy of novel anti-atherosclerosis therapies.

One of the main limitations of the technology is the fact that  individual PET scans are more expensive than ``conventional'' imaging with computed tomography (CT) and magnetic resonance imaging (MRI). Reducing costs of PET scanners is, therefore, a major priority to facilitate the expansion of the technology. Conversely, the combination of PET with CT and MRI often leads to much improved scans, since the structural information offered by CT and MRI can be combined with the functional information offered by PET. 

%In conclusion, the medical applications of PET can be improved by:
%\begin{enumerate}
%\item {\bf Increasing the performance of the PET scanner}, which in turn means improving the energy resolution, spatial resolution, coincidence resolution time and sensitivity, as will be discussed further beyond.
%\item {\bf Reducing the cost of the PET scanner}, which requires the use of detection media which are cheaper than current solid scintillating detectors such as LSO/LYSO without compromising performance.
%\item {\bf Combining PET with structural scans}, which, in the case of MRI requires the use of materials and sensors which can operate in the presence of very intense, highly variable magnetic fields. 
%\end{enumerate}

\section{Performance characteristics of PET scanners}
\label{sec.ppet}

A major goal of PET studies is to obtain a good quality and detailed image of an object. Several parameters associated with the PET scanner are critical to good quality image formation, including energy resolution (ER) spatial resolution (SP), sensitivity, noise, scattered radiations, and contrast. These parameters are interdependent, and if one parameter is improved, one or more of the others may be compromised. We present a brief summary below.

\subsection{True, scattered and random coincidences in a PET detector}

\begin{figure}[!bthp]
	\centering
	\includegraphics[scale=1.0]{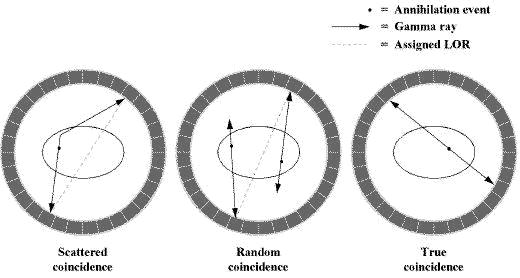}
	\caption{\label{fig.coi} Coincidence detection principle in a PET detector.}
\end{figure}

The reconstruction of the image in a PET system requires crossing many LOR which in turn define one emission point in the area under study. LOR are formed by detecting the coincidence of two photons. Three types of coincidences, illustrated in Figure \ref{fig.coi} are relevant:

\begin{itemize}
\item {\bf True coincidences} occur when both photons from an annihilation event are detected by detectors, neither photon undergoes any form of interaction prior to detection, and no other event is detected within the coincidence time-window.
\item {\bf A scattered coincidence} is one in which one of the detected photons (sometimes both) has undergone at least one Compton scattering event prior to detection. Since the direction of the photon changes due to the scattering process, the resulting coincidence event will be, most likely produce a wrong LOR. Scattered coincidences add a background to the true coincidence distribution which changes slowly with position, decreasing contrast and causing the isotope concentrations to be overestimated. They also add statistical noise to the signal. The number of scattered events detected depends on the volume and attenuation characteristics of the tissue being imaged. The best way to reject scattered coincidences is to build a PET system based on detectors with excellent energy resolution, since the scattered photons have also lost a fraction of their energy, and can therefore be rejected by imposing that the measured energy is inside a narrow window around 511 keV.
\item {\bf Random coincidences} occur when two photons, not arising from the same annihilation event, impinge the detectors within the coincidence time window of the system. As with scattered events, the number of random coincidences detected also depends on the volume and attenuation characteristics of the object being imaged, and on the geometry of the camera. The distribution of random coincidences is fairly uniform across the field of view and will cause isotope concentrations to be overestimated if not corrected for. Random coincidences also add statistical noise to the data. The best way to reject random coincidences is to build a PET system based on detectors with excellent time coincidence resolution (CRT), since in this case one can impose a very narrow coincidence window, therefore minimizing the number of random coincidences which is proportional to CRT. 
\end{itemize}

\subsection{Sensitivity}

The sensitivity of a PET scanner is defined as the number of counts per unit time detected by the device for each unit of activity present in a source. It is normally expressed in counts per second per microcurie (or megabecquerel) (cps/$\mu$Ci or cps/kBq). Assuming that dead time is small, sensitivity depends on the geometric efficiency and detection efficiency. The detection efficiency of a detector depends on the scintillation decay time, density, atomic number, and thickness of the detector material. Thus, for example, Lutetium Oxyorthosilicate (LSO) has a very large density (7.4 g/cc) which results in an attenuation length at 511 keV of 12 mm. LXe is less dense (3 g/cc), and has, therefore, a longer attenuation length, 36 mm at 511 keV. This implies that a LXe detector needs to be 3 times thicker than a LSO detector to achieve the same detection efficiency.

The geometric efficiency of a PET scanner is defined by the solid angle projected by the source of activity at the detector. The geometric factor depends on the distance between the source and the detector, the diameter of the ring and the number of detectors in the ring. Increasing the distance between the detector and the source reduces the solid angle and thus decreases the geometric efficiency of the scanner and vice versa. Increasing the diameter of the ring decreases the solid angle subtended by the source at the detector, thus reducing the geometric efficiency and in turn the sensitivity. Also the sensitivity increases with increasing number of rings in the scanner. This, in turn, requires a cost per detector as low as possible. LXe is 5 times cheaper than LSO (per unit detection, that is, taking into account that the detector must be 3 times thicker in the case of LXe than in the case of LSO) and can be more sparsely instrumented, at least for some applications, thus making it possible the construction of more rings, and therefore increasing the geometrical efficiency and the sensitivity. 

A simple formula for the sensitivity of a PET scanner is:
\begin{equation}
S = \frac{A \cdot \epsilon^2 \cdot e^{-\mu t} \cdot \xi}{4 \pi r^2} (cps/\mu Ci)
\label{eq.sensi}
\end{equation}
where $\xi = 3.7 \times 10^4$~is a numerical conversion factor, $A$~is the detector area seen by a point source to be image, $\epsilon$~is the detector efficiency (e.g, the fraction of the time the detector is alive, which in turn depends on the detector scintillating time and the response of sensors, multiplied by the fraction of events that are relevant for detection, e.g, photoelectric interactions), $\mu$~is the linear attenuation coefficient of 511 keV photons in the detector material, $t$~is the detector thickness. Notice that the factor $\epsilon^2$~comes from the need to form a coincidence with two detectors with efficiency $\epsilon$.

Equation \ref{eq.sensi} is valid for a point source at the center of a single ring scanner. For an extended source at the center of such scanners, it has been shown that the geometric efficiency is approximated as w/2r, where w is the axial width of the detector element and r is the radius of the ring. Thus the sensitivity of a scanner is highest at the center of the axial field-of-view (FOV) and gradually decreases toward the periphery. In typical PET scanners, there are also multiple rings and each detector is connected in coincidence with as many as half the number of detectors on the opposite side in the same ring as well as with detectors in other rings. Thus the sensitivity of multiring scanners will increase with the number of rings.

\subsection{Time of flight (TOF) application in a PET detector}

\begin{figure}[!bthp]
	\centering
	\includegraphics[scale=1.0]{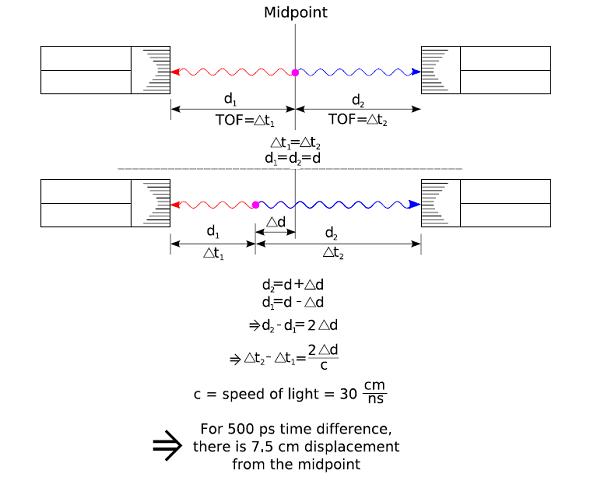}
	\caption{\label{fig.tof} Time of flight (TOF) principle in a PET detector.}
\end{figure}

A Time-of-flight (TOF) PET scanner takes advantage of the difference in arrival times of two photons from the same annihilation event to infer spatial information of this event. 

To understand the principle of TOF applied to PET is important to recall that light travels 30 cm in 1 ns. Consider the situation illustrated in Figure \ref{fig.tof}. In principle one can measure the point of emission along the LOR by taking the time of flight difference between the arrival of the two photons. Since (Figure \ref{fig.tof}):
\begin{equation}
\Delta t = \Delta t_2 - \Delta t_1 = \frac{2 \Delta d}{c}
\end{equation}
and $c = 30 cm/ns$, it follows that the displacement from the mid point
$\Delta d$ is related with the TOF difference between the two photons $\Delta t$ and the speed of light $c$ as:

\begin{equation}
\Delta d =c \frac{\Delta t}{2}
\end{equation}

Thus, if one is able to measure $\Delta t$~ to a resolution of 500 ps (corresponding to the
best time resolution achieved by current commercial systems), the resulting precision in the
determination of $\Delta t$~is 7.5 cm, to be compared with the typical resolution achieved by
conventional PET, which is of the order of 5 mm. A $\Delta t$~resolution of 25 ps would yield a resolution of better than 4 mm, competitive with that achieved in conventional PET scanner, and therefore TOF could be used to determine directly the emission point. 

\begin{figure}[!bhtp]
	\centering
	\includegraphics[scale=0.6]{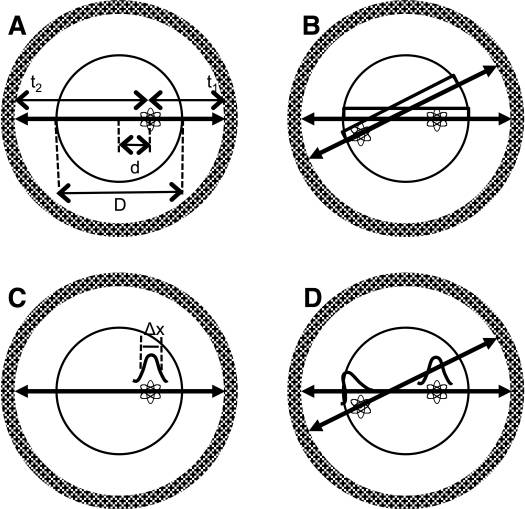}
	\caption{\label{fig.tile} (A) Emission point at a distance $d$ from the center of the scanner within an object of diameter $D$. The two 511 keV photons are detected in coincidence at times $t_1$ and $t_2$. (B) Without precise TOF measurement a uniform probability along the LOR within the object is assumed for each emission point, leading to noise correlations over a portion of image space between the two events as shown in (B). (C) With TOF information the position of the emission point is localized along the LOR with a precision that is defined by a Gaussian distribution of width $\Delta x$. (D) Better localization of the two emission events along their individual LORs leads to reduced noise correlation of the events in image space image reconstruction. }
\end{figure}

In a TOF-PET scanner one uses good time-of-flight resolution to reduce the number of random coincidences, as illustrated in Figure \ref{fig.tile}. The principle is as follows. In a conventional PET without TOF one does not localize the emission point along the LOR. By collecting all possible LORs around the object (full angular coverage) and assuming uniform probability of the emission points lying along the full length of the LORs (and within object boundary), it is mathematically possible to reconstruct the emission object accurately (Figure \ref{fig.tile}, panels A and B). Knowledge of emission point locations along the LORs is not necessary to reconstruct the emission object. However, by assuming uniform probability of event location along the full LOR length, noise from different emission events gets forward and back projected during image reconstruction over many image voxels leading to increased noise correlation. Hence, the image signal-to-noise ratio (SNR) gets reduced.

In time-of-flight (TOF) PET the difference in the arrival times $(t_2 - t_1)$ of the two photons is measured with a precision $\Delta t$~(called coincidence timing resolution, CRT) that helps localize the emission point along the LOR within a small region of the object, as illustrated in  Figure \ref{fig.tile}, panel C. The uncertainty in this localization is determined by the CRT. The corresponding uncertainty in spatial localization $\Delta x$~ along the LOR is given by 
$\Delta x=c \times \Delta t/2$. As previously noted, if $\Delta x$~ is the same or smaller than the detector spatial resolution then in principle image reconstruction is not needed. Typically this spatial localization, however, is more than one order of magnitude worse than the detector spatial resolution, and hence image reconstruction is still necessary to produce tomographic images. However, during reconstruction, noise from different events is now forward and back projected over only a limited number of image voxels as defined by the spatial uncertainty, leading to reduced noise correlations and improved image signal-to-noise ratio (SNR) as illustrated in  Figure \ref{fig.tile}, panel D.

\begin{figure}[!bhtp]
	\centering
	\includegraphics[scale=0.5]{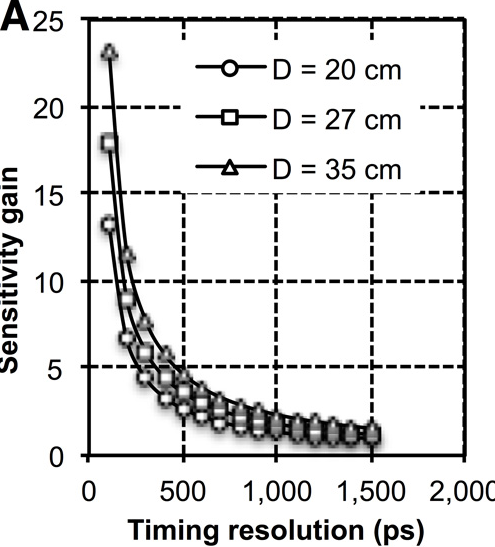}
	\caption{\label{fig.sensi} Gain in sensitivity (defined in the text) as defined by $D/\Delta x$ plotted as a function of timing resolution for cylindrical ``phantoms'' with three different diameters.}
\end{figure}

With the knowledge that during the forward and back projection steps in image reconstruction noise will be spread over fewer voxels along the LOR (defined by $\Delta x$), it has been  shown  that the effective gain in sensitivity at the center of a uniform cylinder due to TOF information is given by $D/\Delta x$ \cite{snyder81,budinger83}. Figure \ref{fig.sensi} shows a plot of this gain in sensitivity plotted as a function of timing resolution and for varying object sizes. As the object size increases or timing resolution improves the gain due to TOF PET increases. 

Consider as an example that one is performing a torso scan, ($D \sim 30$~cm). A PET capable of a CRT of 200 ps will result in $\Delta x = 3$~ cm and thus the gain in sensitivity may be as high as $30/3 = 10$). 
  
It follows that a PET scanner capable of a CRT in the range of 100 ps can improve the sensitivity by roughly a factor 10 with respect to a conventional PET with no TOF measurement.

\subsection{Summary: main requirements for a PET scanner}

From the discussion above we can conclude that the 
main requirements for a PET scanner are: 

\begin{enumerate}
\item {\bf Dense detectors}, with high stopping power for 511 keV gammas.
\item {\bf Spatial resolution of sub mm to few mm}, depending on the application (small animal PET requires sub mm resolution while full body PET can be done with a resolution of few mm).
\item {\bf Energy resolution as good as possible}, to eliminate Compton coincidences.
\item {\bf High count rate capability ($\sim10^6$~ s$^{-1}$ per cm$^2$~ of detecting surface}, which results in minimizing scan times and/or doses.
\item {\bf Fast scintillating time}, which allows narrow CRT and possibilities TOF application. 
\item {\bf Large angular acceptance}, for torso, ``full body PET'', which in turn requires a large axial (along the patient's body) coverage. This requires cheap detection systems. 
\end{enumerate}

\section{Solid scintillation detectors versus LXe}
\label{sec.ssd}

\subsection{Solid scintillation detectors}

Today's conventional PETs use solid scintillation detectors (SSD) such as Sodium Iodine (NaI), Bismuth Germanate (BGO) or Lutetium oxyorthosilicate (LSO), and Lutetium-yttrium oxyorthosilicate (LYSO) readout by light sensitive detectors. The standard has shifted from the use of NaI crystals to the use of BGO and LSO devices. Until recently, the SSDs were readout with photomultipliers (PMTs), but the so-called Silicon Photomultipliers (SiPMs) are emerging in the last few years as a major alternative. Unlike PMTs, SiPMs can be used in the presence of a strong magnetic field, therefore opening the possibility to build Nuclear Magnetic Resonance (MRI) compatible PET devices. 

The physical properties that define SSDs are: 
\begin{enumerate}
\item {\bf Attenuation length ($\lambda$)}, which sets the scale of the length (across the photon line of flight) that the detector has to have in order to stop most of the incoming radiation.
\item {\bf Density ($\rho$)}, which is related with the total size and weight of the detector.
\item {\bf Photon yield per keV ($Y$)}, which must be as high as possible to record large signals. 
\item {\bf Energy resolution at 511 keV ($\sigma_E$)}, which must be as good as possible. 
\item {\bf Transverse spatial resolution ($\sigma_T$)} (relative to the photon line of flight), which in turn depends on the photon yield and the granularity of the readout sensors.
\item {\bf Longitudinal spatial resolution ($\sigma_L$)}, important to minimize the so-called parallax error.  $\sigma_L$~  tends to be poor for the SSDs, which do not measure the longitudinal coordinate (thus $\sigma_L \sim L/\sqrt{12}$, where $L$~is the detector length). 
\item {\bf Scintillation decay time ($t_s$)}, which must be as fast as possible, to maximize the number of events acquired per unit time and to minimize the window used to correlate events in different crystals. In addition, if the system has very good time resolution (in the range of few hundred picoseconds) time-of-flight measurements (TOF) are possible.  
\end{enumerate}

\subsection{Liquid Xenon as detection material}

%
%\begin{figure}[!bhtp]
%	\centering
%	\includegraphics[scale=0.5]{img/XeLight.pdf}
%	\caption{\label{fig.xel} Scintillation and ionization properties of LXe.}
%\end{figure}

Xenon is a noble gas. It responds to ionizing radiation providing both ionization and scintillation signals. The ionization signal is due to atomic electrons ejected from the xenon atoms by the incoming radiation, which take a long time to recombine due to the noble-gas nature of xenon (and therefore can be drifted to a collection electrode, if so desired). The scintillation signal is due to the de-excitation of xenon atoms forming dimers which decay after 2.2 ns (dominant single mode) or 27 ns (triplet mode) emitting ultraviolet light (VUV) of 178 nm wavelength %(Figure \ref{fig.xel}).

If $E$~ is the energy deposited by the ionizing radiation (in this case 511 keV), the maximum scintillation yield of LXe is given as $E/W_{ph}$, where $W_{ph}$~ is the average energy required for the production of a single photon. The most probable value of $W_{ph}$~ in LXe  is  $13.8 \pm 0.9$~eV \cite{aprile10}. Therefore, a maximum of 37,000 scintillation photons are produced when a 511 keV gamma interacts in the LXe. On the other hand, measurements carried out with electrons of 1 MeV result in a value of $W_{ph}^e = 21.6$ \cite{doke02}. This value is attributed to ionization electrons that do not recombine, and would imply a yield of $\sim$ 24,000 scintillation photons for a 511 keV gammas. The results presented in this work are obtained assuming the maximum scintillation yield, but we also discuss the implications of a lower scintillation yield in the energy and spacial resolution. 

In its liquid phase (at a temperature of 165 K and 1 bar of pressure) LXe has a reasonable high density (3 g/cc) and an acceptable attenuation length for 511 keV gammas (36 mm), which makes it suitable for PET applications. Its advantage with respect to SSD are: a) its very high yield (24,000--37,000 photons for a 511 keV gamma), which in turn can translate in excellent energy and transverse spatial resolution; b) Its ability to provide a 3D measurement of the interaction point and thus a high-resolution measurement of the longitudinal coordinate, minimizing parallax errors; c) its capability to identify Compton events depositing all its energy in the detector as separate-site interaction, due to the relatively large interaction length in xenon;  d) its very fast scintillation decay time, which makes it suitable as a TOF-PET;  and e) its relatively low cost (e.g, about 10\% of the cost of LSO per unit detector).  Table \ref{table.SDPP} shows the physical properties of common PET SSD compared with that of liquid xenon (LXe).

\begin{table}[htdp!]
\caption{Physical properties of common PET SSD and of LXe}
\begin{center}
\begin{tabular}{l|cccc}
\toprule
& \textbf{NaI} & \textbf{BGO} & \textbf{LSO} & \textbf{LXe}\\
\hline
Effective $Z$ & 50 & 74 & 66 & 54 \\
$\rho$~(g/cm$^3$) & 3.7 & 7.1 & 7.4 &  3 \\
$\lambda$ ~ at 511 keV (mm) & 28 & 11 & 12 & 36 \\
$Y$~per keV & 38 & 6 & 29 & {\bf 72.4} \\
%RLO & 100 & 15 & 75 & {\bf 190} \\
$t_s$~(ns) & 230 & 300 & 40 & {\bf 2.2} \\
\toprule
\end{tabular}
\end{center}
\label{table.SDPP}
\end{table}%

\section{The PETALO concept}
\label{sec.petalo}

The first idea of using a Liquid Xenon Time Projection Chamber (LXeTPC) for PET was proposed in 1993 by Chepel \cite{chepel02}. The proposed detector was a LXe multi-wire detector consisting of six ionization cells, each formed by two parallel cathode plates with a multi-wire anode in the middle. 
Subsequent R\&D is documented in \cite{chepel94,chepel95,lopes95,chepel97,crespo98,chepel99,crespo00}. {\bf All those devices were based in the exploration of the ionization signal in LXe}. {\em However, the measurement of such signals introduces a severe constrain to the technique, given the slow drift time of electrons in LXe} (typically of the order of 2 mm/$\mu$s, for a drifting field of 1 KV/cm). Since $\lambda = 3$.6~cm the practical length of a LXe cell (along the photon line of flight) must be of 5 cm to contain 77 \% of the photons. This, in turn, implies drifting times of 25 $\mu$s, which limits the interaction rate that can be recorded by the cell to $\sim10^5 s^{-1}$, and therefore imposes a low-rate PET with a limited range of applications. 

On the other hand, the possibility of building a LXe PET (with TOF capabilities) based on the excellent properties of LXe as scintillator, was first suggested by Lavoie in 1976 \cite{lavoie76} , and the study of this type of PET was carried out by the Waseda group \cite{doke06,nishikido05,nishikido04}. The Waseda prototype was based in LXe cells read out by VUV-sensitive PMTs. In those cells one of the sides was left instrumented. The relatively poor performance of the system can be attributed in part to the use of PMTs and in part to the partial lack of instrumentation which affected both the energy and the time resolution. The PMTs, although sensitive to the VUV light emitted by xenon had low quantum efficiencies (in the range 5-25 \%), and their rather large size compared with the size of the cell ($18\times 18$~mm$^2$) introduced significant geometrical effects which were difficult to correct. As a result of the above effects combined the energy resolution was of the same order than that of conventional SSDs. The space resolution was rather good, in the range of 2-3 mm, but only in the central volume of the cell (a cube of 5 mm size), and deteriorated rapidly in the borders, due to the space corrections introduced by the (relatively) large PMTs. The time resolution in the central volume was excellent, of the order of 260 ps, showing the enormous potential of the technology for PET application.

%\begin{figure}[!htb]
%	\begin{center}
%           \subfloat[]{
%               \includegraphics[width=.5\textwidth]{img/waseda1.png}
%           }
%           \subfloat[]{
%               \includegraphics[width=.5\textwidth]{img/waseda2.png}
%           }
%           \caption{\label{fig.waseda} Illustration of Waseda's group prototype taken from \cite{nishikido05}. (a) Cross-sectional view of the prototype liquid Xe PET detector. (b) Arrangement of 32 PMTs.}
%    \end{center}
%\end{figure}

PETALO (Positron Emission TOF Apparatus based on Liquid xenOn) is a new concept for a TOF-capable, high sensitivity, PET apparatus based in the excellent scintillating properties of LXe and the concept of a new type of detection cell, which captures with high efficiency, minimal border effects and uniform response, most of the light produced.
 
\begin{itemize} 
\item {\bf High efficiency} is achieved by covering the internal side of the cell with reflective panels, made of high density teflon coated with TPB. The emitted ultraviolet light (172 nm) is shifted to 420 nm as soon as it hits the teflon panels, which on the other hand, reflect blue light in the range of 420 nm with 98-99\% efficiency. Furthermore, TPB does not absorb blue light above 400 nm, therefore minimising loses. 
\item {\bf High homogeneity and uniform response}, with minimal border effects is achieved by choosing SiPMs as readout devices. Currently, several manufacturers offer SiPMs of large area, high gain, low dark current and very low noise which can operate at liquid xenon temperatures with excellent performance.  
\item {\bf Excellent energy resolution} is achieved thanks to the hight light yield of LXe and the homogeneity of the cell (the measured energy varies very little from one point of the cell to the other).  
\item {\bf Good spatial resolution in the three coordinates} is possible by using the SiPMs array covering the entry and exit face of the cell with SiPMs which provide the transverse (x-y) coordinates and using the ratio of light recorded in the entry and exit face to measure the longitudinal (z) coordinate. 
\item {\bf Excellent CRT} is possible thanks to the fast decay scintillation time of LXe (2.2 ns) and the fast response of the SiPMs.  
\end{itemize}

\section{Silicon Photomultipliers}
\label{sec.sipm}

When a photon travels through silicon (a semiconductor), it can transfer its energy to a bound state (valence) electron, pushing it into the conduction band and thus creating an electron-hole pair. Silicon is a good photo detector material in the spectral range form 350 nm up to 800 nm, e.g, at frequencies ranging from the deep blue to the red. 

In a silicon photodiode one applies a reverse bias to a p-n junction. A reverse bias is simply a voltage that counteracts the diffusive force that draws negative and positive charge carriers (electrons and holes respectively) to the depletion region situated around the p-n junction. Under application of the reverse bias, the diode is ``open'' (no flow of net charge) until it absorbs a photon. When this happens, electrons acquire enough energy to be pushed to the conduction band and as a  result a net current (electrons through the n-type and holes through the p-type sides of the device) appears in the diode.

When a sufficiently high electric field ($>5 \times 10^5$~ V/cm) is generated within the depletion region of the silicon, a charge carrier created in this region will be accelerated to a point where it carries sufficient kinetic energy to create secondary charge pairs through a process called impact ionization. In this way, a single photoelectron can trigger a self-perpetuating ionization cascade that will spread throughout the silicon volume subjected to the field. The silicon will break down and become conductive, effectively amplifying the original photoelectron into a macroscopic current flow. This process is called Geiger discharge, in analogy to the ionization discharge observed in a Geiger-Muller tube. 

A photodiode operated in Geiger mode employs this mechanism of breakdown to achieve a high gain. The p-n junction region is designed in such a way that it can sustain a reverse bias beyond its nominal breakdown voltage, creating the necessary high field gradients across the junction. Once a current is flowing it should then be stopped or ‘quenched’. Using passive quenching (i.e. no active circuitry), this is achieved through the means of a series resistor RQ which limits the current drawn by the diode during break down, and hence lowers the reverse voltage seen by the diode to a value below its breakdown voltage. Thus a cycle of breakdown, avalanche, quench and subsequent reset of the bias to a value above the breakdown voltage is achieved. In this way, a single photodiode device operated in Geiger-mode functions as a photon-triggered switch, in either an ``on'' or ``off'' state, and therefore cannot provide proportional information regarding the magnitude of an instantaneous photon flux. Regardless of the number of photons interacting within a diode at the same time, it will produce a signal of ``1'' photon.

\begin{figure}[!bhtp]
	\centering
	\includegraphics[scale=1.1]{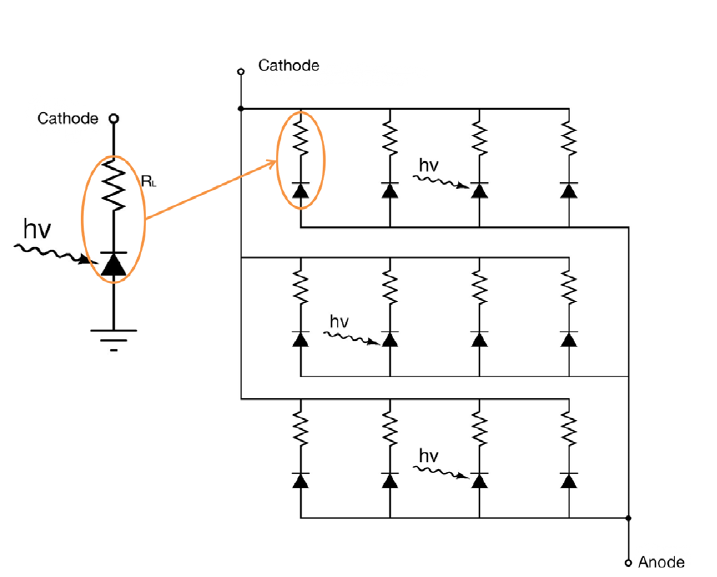}
	\includegraphics[scale=1.1]{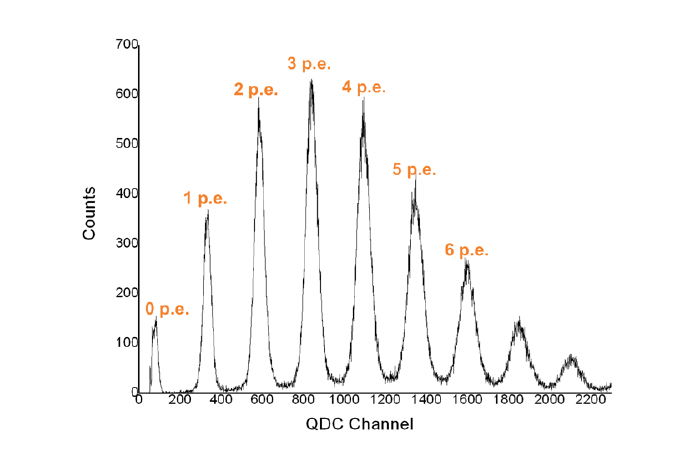}
	\caption{\label{fig.sipmc} Left: an array of microcells (photodiode plus quenching resistor) with their output summed up. Right: photoelectron spectrum of the SiPM, achieved using brief, low-level light pulses.}
\end{figure}

To overcome this lack of proportionality, the Silicon Photomultiplier integrates a dense array of small, electrically and optically isolated Geiger-mode photodiodes. Each photodiode element in the array is referred to as a ``microcell''. Depending on the application there can be between 100 and 1000 microcells per mm$^2$, each one with  its own quenching resistor. The signals of all microcells are then summed to form the output of the SiPM. A simplified electric circuit to illustrate the concept is shown in Figure \ref{fig.sipmc} (left panel). Each microcell detects photons identically and independently. The sum of the discharge currents from each of these individual binary detectors combines to form a quasi-analog output, and is thus capable of giving information on the magnitude of an incident photon flux. The response to low-level light pulses is shown in in Figure \ref{fig.sipmc} (right panel).

\subsection{Silicon photomultipliers performance parameters}
The the main parameters that characterize a SiPM are: gain, PDE, noise, dynamic range, timing and temperature sensitivity. We briefly discuss those parameters here.

\subsubsection*{Over-Voltage}
\begin{figure}[!bhtp]
	\centering
	\includegraphics[scale=0.9]{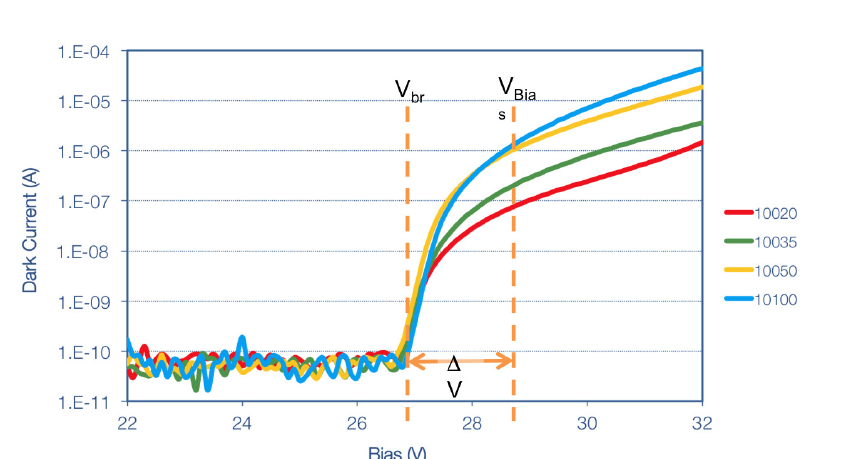}
	\includegraphics[scale=0.9]{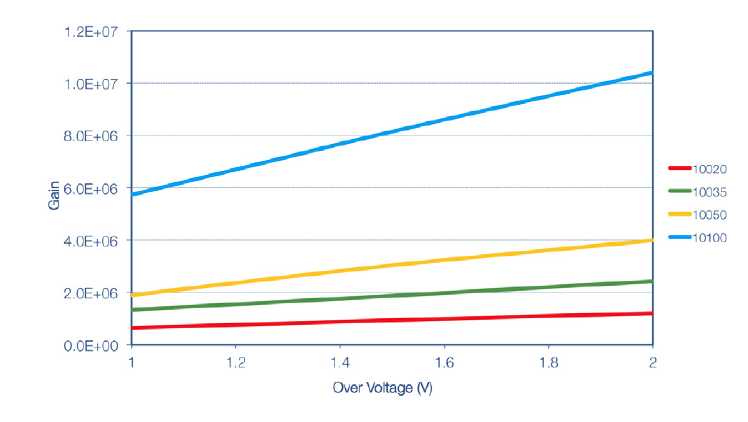}
	\caption{\label{fig.sipp1} Left: Dark current as a function of the voltage for 1 mm SiPMs of various microcell size (20 $\mu$m, 35 $\mu$m, 50 $\mu$m, 100 $\mu$m). Right: gain as a function of the over-voltage for various microcell size (20 $\mu$m, 35 $\mu$m, 50 $\mu$m, 100 $\mu$m).}
\end{figure}

The breakdown voltage (V$_{br}$) is the bias point at which the electric field strength generated in the depletion region is sufficient to create a Geiger discharge. The point of breakdown is clearly seen on an I-V plot by the sudden increase in current, as in 
Figure \ref{fig.sipp1} (left panel). The bias voltage (V$_{bias}$) is typically set at about 2V above the breakdown voltage. This 2V gap is referred to as the ``over-voltage'' 
($\Delta$V) and is critical in defining the important performance parameters of the SiPM.

\subsubsection*{Gain}

Each microcell in an SiPM is comprised of a Geiger-mode photodiode in series with an integrated quench resistor. Each microcell generates a highly uniform and quantized amount of charge every time the microcell undergoes a Geiger breakdown. The gain of a microcell (and hence the detector) is defined as the ratio of the output charge to the charge on an electron. The output charge can be calculated from the over-voltage and the microcell capacitance.

\begin{equation}
G = \frac{C \cdot \Delta V}{q}
\end{equation}

Figure \ref{fig.sipp1} (right panel) shows the gain as a function of the over-voltage  for
different microcell sizes. 

\subsubsection*{Photon Detection Efficiency (PDE)}

\begin{figure}[!bhtp]
	\centering
	\includegraphics[scale=0.35]{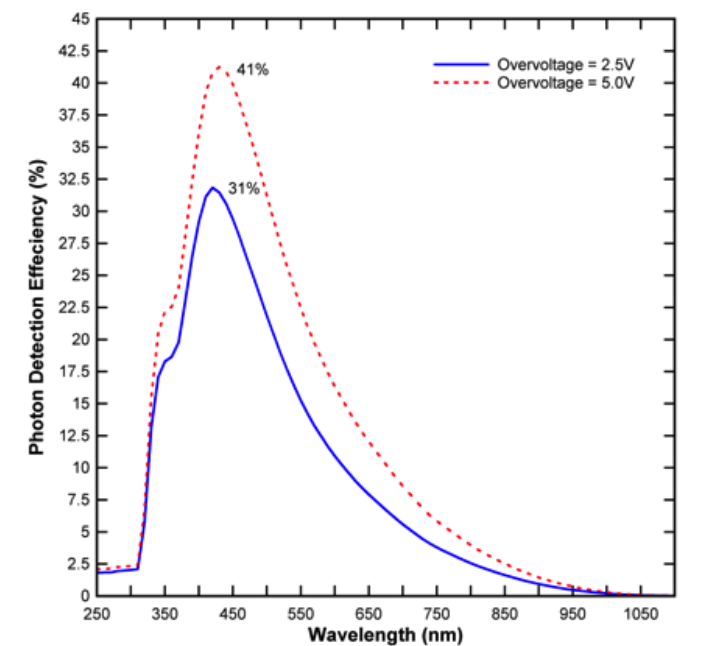}

	\caption{\label{fig.pde} PDE as a function of the wavelength for two different over-voltages.}
\end{figure}

The photon detection efficiency (PDE) of a SiPM is the statistical probability that an incident photon will produce a Geiger pulse from one of the SiPM microcells. It differs slightly from the quantum efficiency (QE) that is quoted for a PMT or APD, due to the microcell structure of the device. The PDE is a function of wavelength and bias and is given by

\begin{equation}
PDE(\lambda, V) = \eta{\lambda} \cdot \epsilon(V) \cdot F
\end{equation}
where $ \eta{\lambda}$~is the quantum efficiency of silicon, $\epsilon(V)$~is the avalanche initiation probability and F is the fill factor of the device. The avalanche initiation probability takes into account the fact that no all generated photoelectrons will initiate an avalanche. The fill factor is the ratio of active to inactive area in the SiPM as a result of the gaps between microcells. The PDE of a typical SiPM is shown in Figure \ref{fig.pde} as a function of the wavelength for for two different over-voltages. The sensitivity of SiPMs peaks around 420 nm with a PDE nearing 45 \% for sufficiently high over-voltage. 

\subsubsection*{Noise}

\begin{figure}[!bhtp]
	\centering
	\includegraphics[scale=0.9]{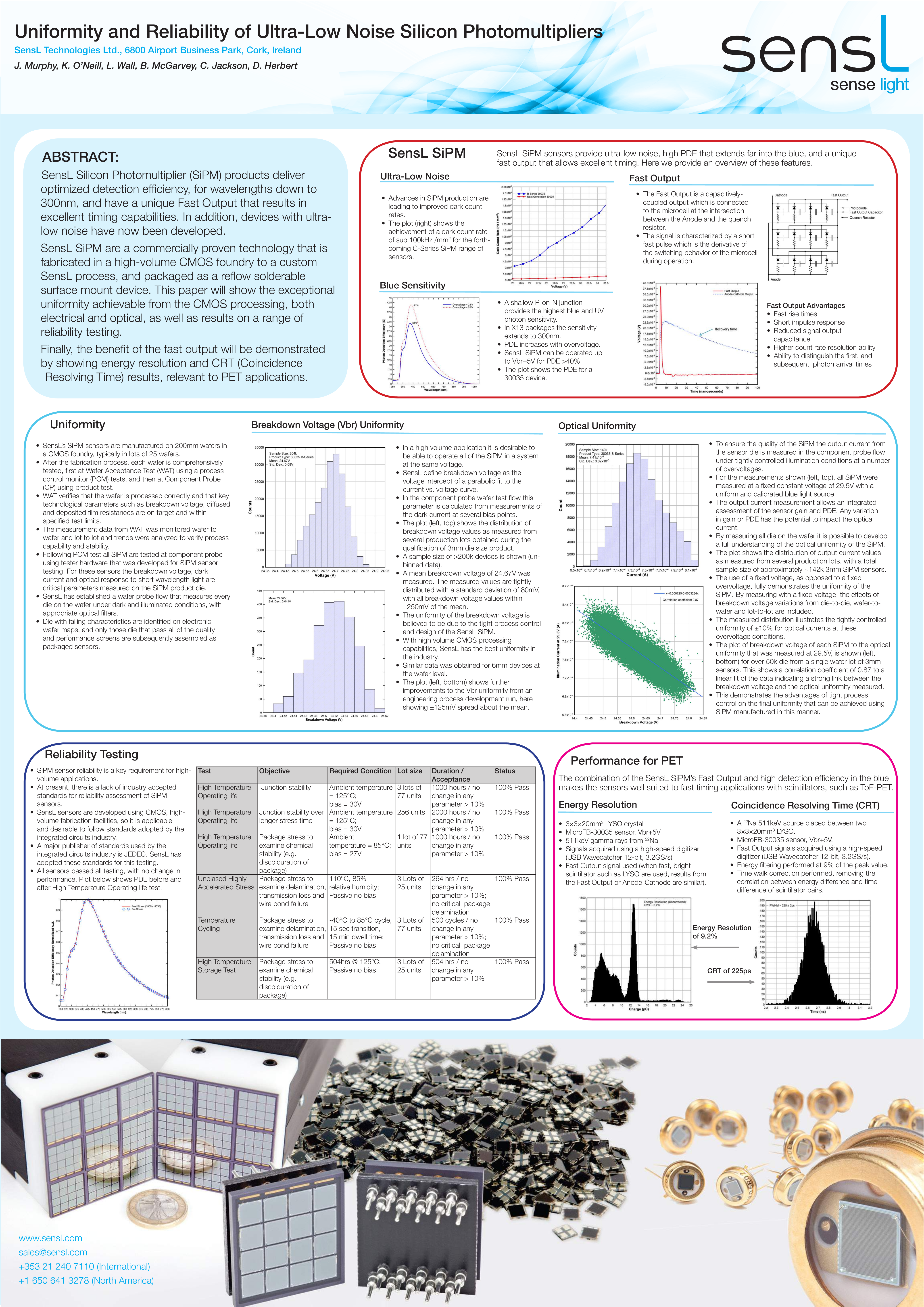}
	\caption{\label{fig.dcr} DCR as a function of the voltage for SENSL B series (blue line) and the new D series (red line) to be used in PETALO. Advances in SiPM manufacturing technology have reduced the DCR in more than one order of magnitude.}
\end{figure}

The main source of noise in an SiPM is the dark count rate (DCR), which is primarily due to thermally generated electrons that create an avalanche in the high field region. The signals resulting from the breakdown of the cell, due to either photoelectrons or thermally generated electrons, are identical. Therefore, these electrons form a source of noise at the single photon level. If a threshold can be set above the single photon level, false triggers from the noise can be avoided, but the dark counts will always form a contribution to the measured signal.
Since this noise is comprised of a series of pulses, its magnitude
is often quoted as a pulse rate, typically in kHz or MHz. Figure \ref{fig.dcr} shows the DCR as a function of the voltage for SENSL B series (blue line) and the new D series (red line) to be used in PETALO. Advances in SiPM manufacturing technology have reduced the DCR in more than one order of magnitude. 

It should be noted that the magnitude of the DCR itself is not the noise contribution. If a given source of noise was always constant, then it could easily be subtracted from the signal. It is instead the fluctuations on the noise which degrade a measurement. The occurrence of the dark pulses is Poissonian in time and so the noise contribution can be taken as the square root of the DCR.

An additional component of SiPM noise is that of optical cross-talk between microcells. When undergoing avalanche, carriers near the junction emit photons as they are accelerated by the high electric field. These photons tend to be in the near infrared region and can travel substantial distances through the device. Typically $2 \times 10^5$ are emitted per electron crossing the junction. These photons can travel to neighboring microcells and may initiate a subsequent Geiger avalanche there. The crosstalk probability is the probability that an avalanching microcell will initiate an avalanche in a second microcell. The process happens instantaneously and as a consequence, single photons may generate signals equivalent to a 2, 3 or higher photoelectron event. The optical crosstalk probability is a function of SiPM over-voltage and the distance between neighboring microcells, and can be estimated by the ratio of the count rate at the second photoelectron level to the count rate at the single photoelectron level. 

\subsubsection*{Temperature dependence}

The most important effect of temperature on the SiPM are a change in the breakdown voltage of the diode and in the dark count rate.

\begin{figure}[!bhtp]
	\centering
	\includegraphics[scale=0.9]{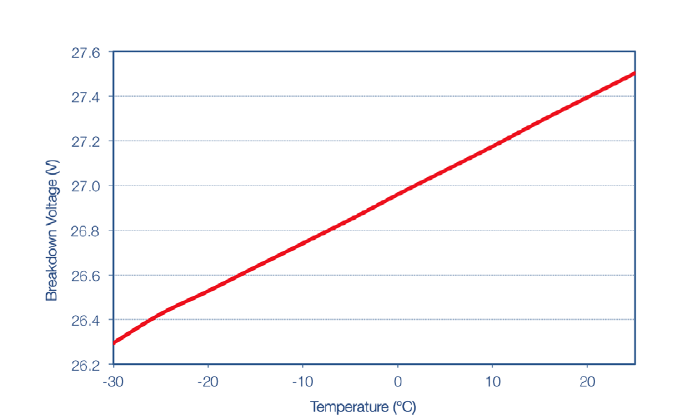}
	\includegraphics[scale=0.9]{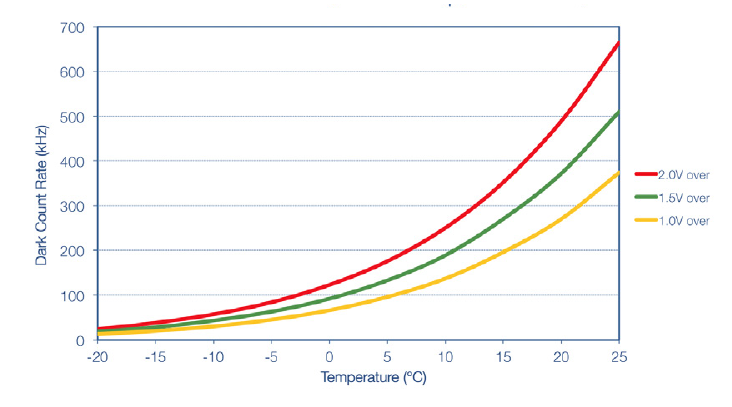}
	\caption{\label{fig.temp} Left: breakdown voltage as a function of the temperature. Right: DCR as a function of the temperature and over-voltage.}
\end{figure}

The breakdown voltage changes as a function of temperature, as shown in Figure \ref{fig.temp} (left panel). If a constant over-voltage is maintained, many parameters, such as gain, PDE and timing, will remain the same as at room temperature. However, regardless of constant over-voltage, the DCR will be altered by a change in temperature, as shown in Figure \ref{fig.temp} (right panel). An increase in temperature will increase dark rate, and the converse is also true: for every 10$^\circ$ C reduction in device temperature, the dark count rate by a factor 2.

\section{The Liquid Xenon Scintillating Cell (LXSC)}
\label{sec.lxsc}

\begin{figure}[!htb]
	\centering
	
	\includegraphics[scale=0.7]{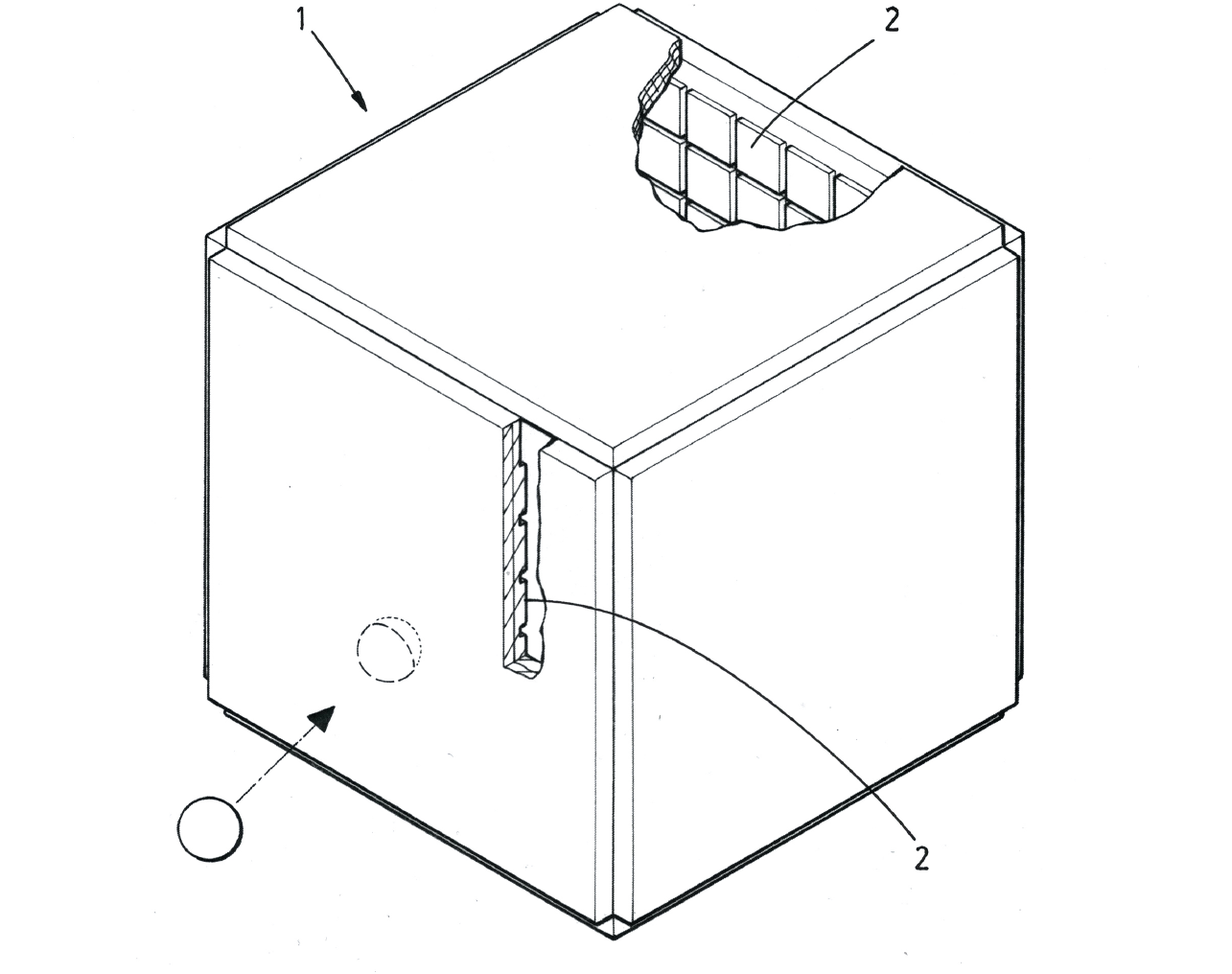}
	\caption{\label{fig.box} The default design of the LXSC (called LXSC2) instruments the entry and exit faces of the box with large silicon photomultipliers (SiPMs), coated with a wavelength shifter, tetraphenyl butadiene (TPB). The non-instrumented faces are covered by reflecting Teflon coated with TPB.  }
\end{figure}

Figure \ref{fig.box}~shows a conceptual drawing of the keystone of the PETALO apparatus, the Liquid Xenon Scintillating Cell (LXSC). The cell is a box  filled with LXe whose transverse dimensions are chosen to optimize packing and with a thickness optimized to contain a large fraction of the incoming photons. The entry and exit faces of the box (relative to the incoming gammas direction) are instrumented with large silicon photomultipliers (SiPMs), coated with a wavelength shifter, tetraphenyl butadiene (TPB). The non-instrumented faces are covered by reflecting Teflon coated with TPB. 

%The transverse dimensions depend on the application (e.g, the diameter of the PET scanner). The default chosen for this study is of $5\times 5$ cm$^2$. The longitudinal dimension is chosen so that most of the incoming 511 keV photons interact in the volume. The default dimension is 5 cm, which results in 77\% of the initial gammas interacting in the cell. 

%The best performance from the point of view of energy resolution is obtained when all the box faces are instrumented with SiPMs, as illustrated in the left panel of Figure \ref{fig.box}. We denote this configuration as LXSC6. On the other hand, the most economical configuration is obtained instrumenting only the entry and exit faces (relative to the beam direction). We call this configuration LXSC2. Other possible configurations instrument three faces (entry, exit and one of the lateral faces, LXSC3) or four faces (entry, exit and two opposite laterals, LXSC4). 
%The faces non instrumented are covered by reflecting teflon coated with TPB. In section \ref{sec.mc} we show that an excellent performance can be obtained even with the sparse LXSC2 configuration. 

%\newpage
%\thispagestyle{newstyle}
%\section{The performance of the LXSC}
%\label{sec.mc}
%

\subsection{Dice Boards}
\label{sec.dc}

In the LXSC the SiPMs are arranged into a {\em Dice Board}, forming a matrix.
The NEXT collaboration \cite{next} has developed DBs \cite{next13,next12} for the NEXT experiment \cite{jj14}, in particular for the tracking plane of the DEMO, NEW and NEXT-100 detectors. 

\begin{figure}[!htb]
	\centering
	\subfloat[Cuflon DB]{
		\includegraphics[width=.5\textwidth]{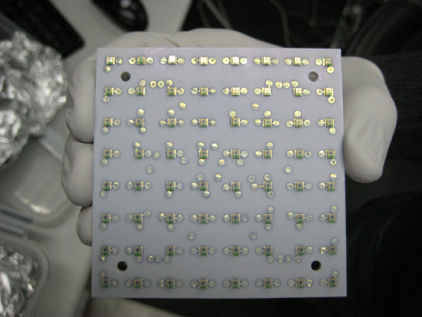}
		\label{fig.db1}
	}
	\subfloat[DB response to UV light]{
    	\includegraphics[width=.5\textwidth]{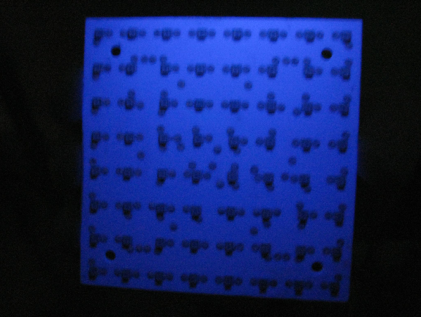}
		\label{fig.db2}
	}
	\caption{\label{fig.DB} Dice boards developed for NEXT-DEMO experiment. (a) DB made of Cuflon and coated with TPB. (b) Response of a DB (emitting blue light) when illuminated with a UV lamp.}
\end{figure}

Figure \ref{fig.db1} shows a DB made of Cuflon, developed for the NEXT-DEMO detector. The DB is coated with TPB, which shifts the VUV light emitted by xenon (170 nm) to blue (420 nm). Figure \ref{fig.db2} shows the response of a DB (emitting blue light) when illuminated with a UV lamp. The DBs of PETALO will have a similar design, except for the use of larger SiPMs (we are currently planning to use 6mm parts, available from several vendors).
Indeed the DBs for PETALO are easier to fabricate and much more economical than those developed for NEXT, since there are no radio purity restrictions, allowing the use of standard PCB materials such as FR4, which are forbidden in a radio pure application.

\section{Performance of the LXSC}
\label{sec.mc}

\subsection{Monte Carlo simulation of the LXSC}

\begin{figure}[!htb]
	\centering
	\subfloat[Photoelectric event]{
		\includegraphics[width=.5\textwidth]{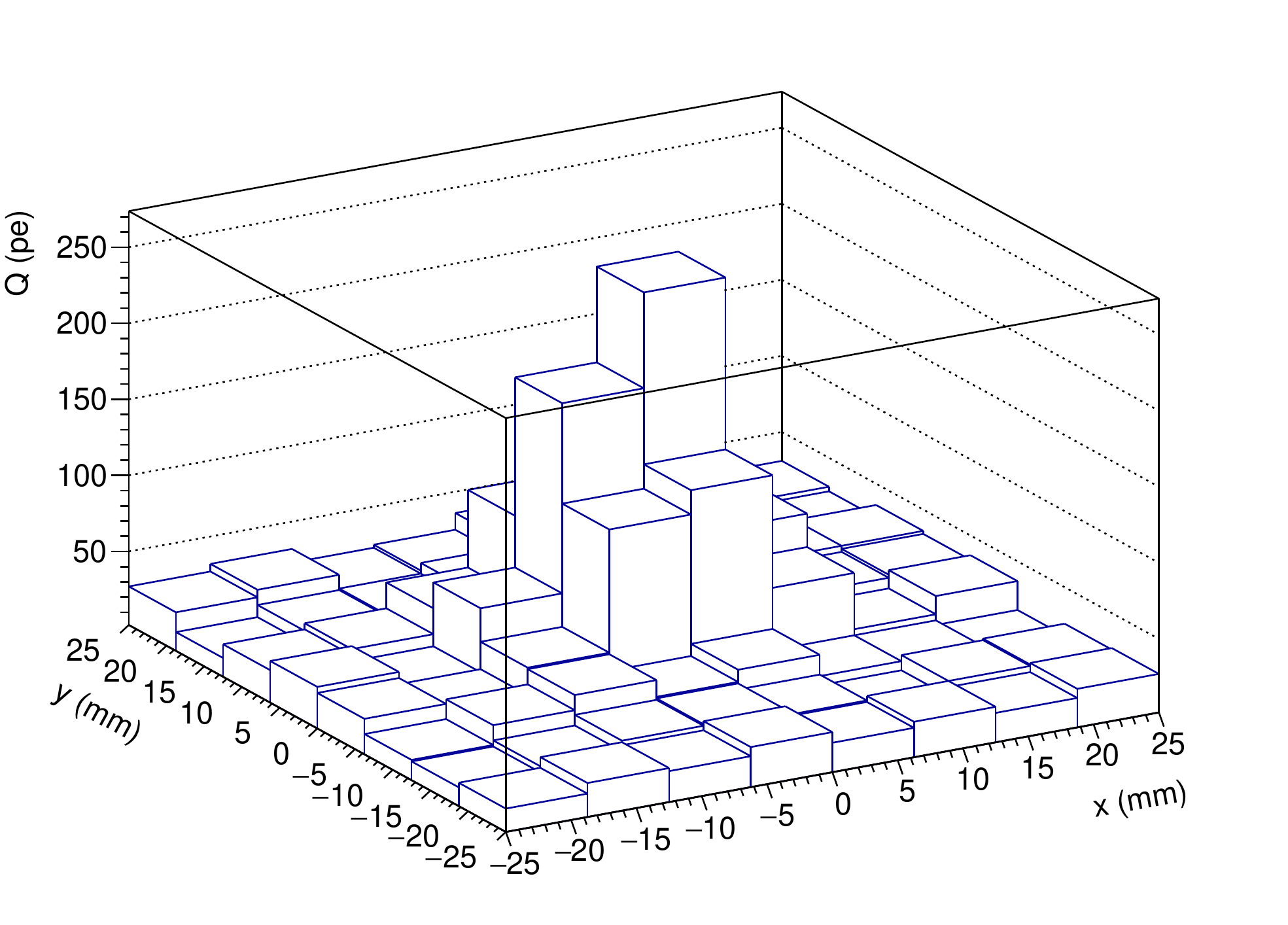}
		\label{fig.photoEvent}
	}
	\subfloat[Compton event]{
		\includegraphics[width=.5\textwidth]{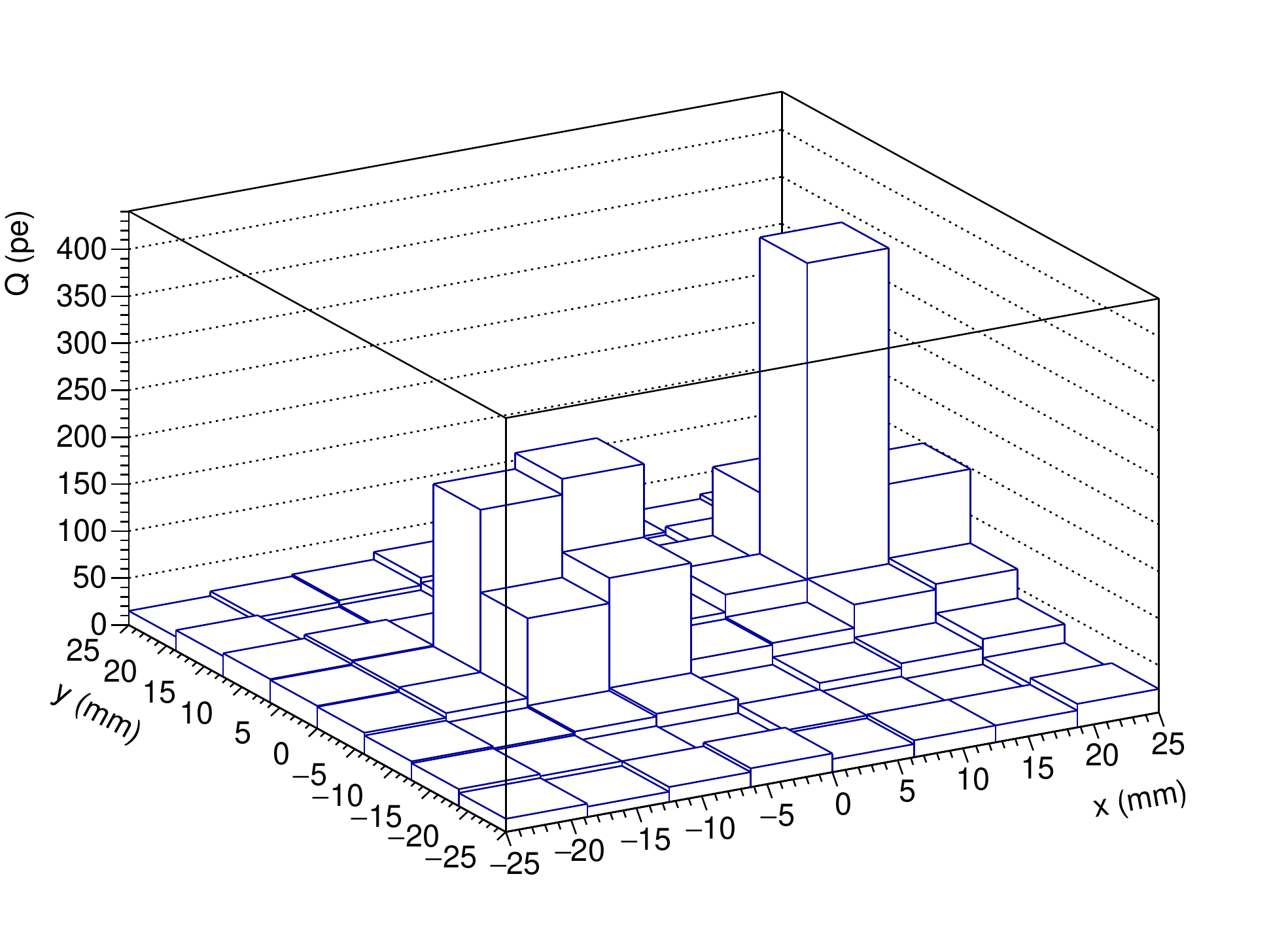}
		\label{fig.comptonEvent}
	}
	\caption{\label{fig.events}  Charge on entry plane for two events in the LXSC. (a) Photoelectric event showing a single deposition cluster. (b) Two-site Compton event.}
\end{figure}

\begin{figure}[!bhtp]
	\centering
	\includegraphics[scale=0.8]{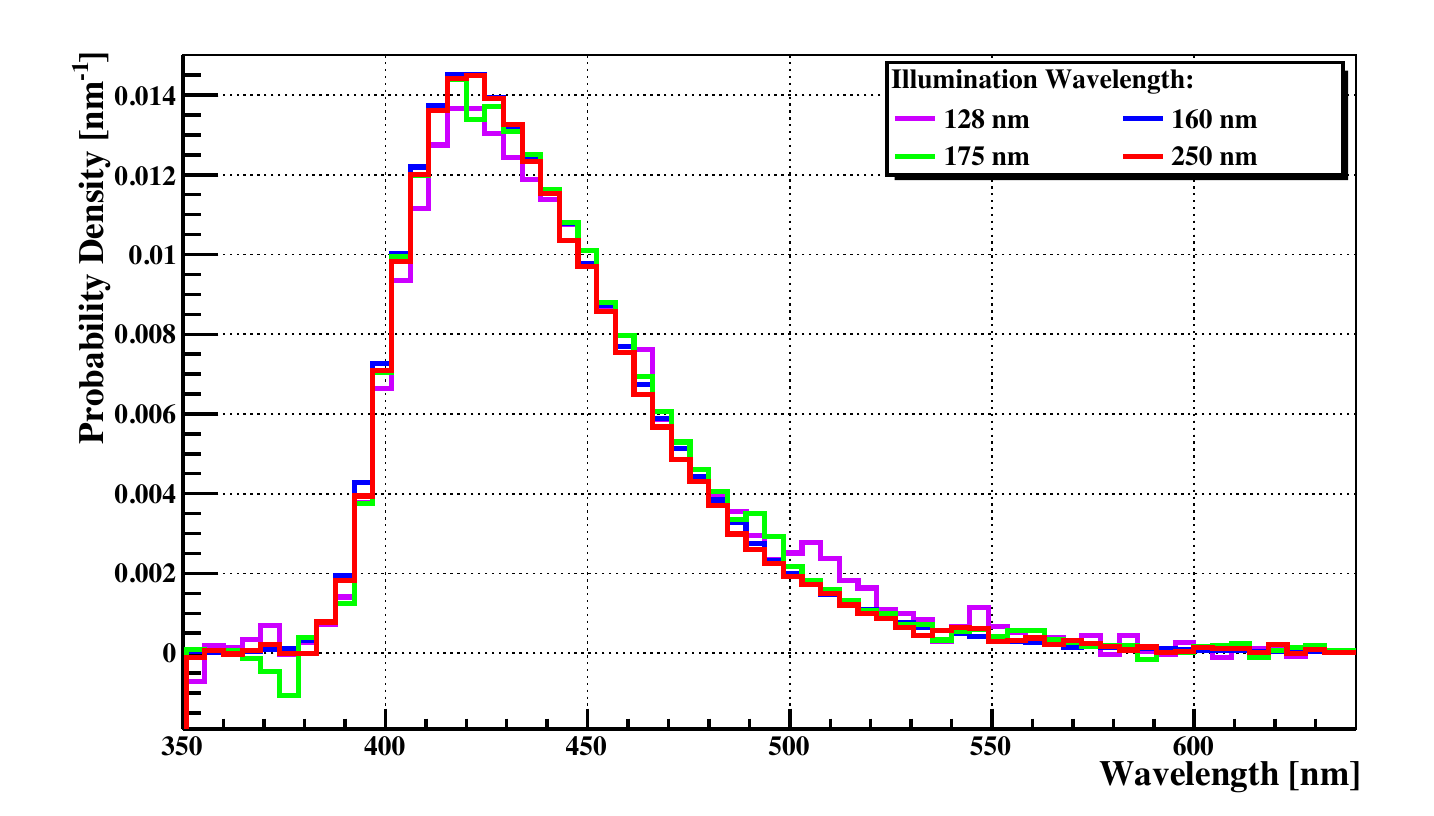}
	
	\caption{\label{fig.tpb} Visible re-emission spectrum for a TPB film illuminated with 128, 160, 175, and 250 nm light. All spectra are normalized to unit area.}
\end{figure}

A full GEANT4 simulation has been carried out to study the performance of different configurations of the LXSC. Gammas of 511 keV enter the LXSC from outside (defining the ``entry face'') and interact in the cell, through photoelectric (around 20\% of the times, see Figure \ref{fig.photoEvent}) or Compton interactions (Figure \ref{fig.comptonEvent}). The produced VUV (172 nm) photons are propagated until they hit one of the surfaces of the box. At his point, the effect of the TPB is taken into account. The VUV is absorbed and 80\% of the times (the probability of re-emission, measured by NEXT and other collaborations) a blue photon is emitted isotropically according to the TPB re-emission spectrum measured by the NEXT collaboration and shown in Figure \ref{fig.tpb}.
\newpage
The emitted photon can then impinge on a SiPM, in which case a signal is generated in the SiPM with a probability following the PDE probability distribution shown in Figure \ref{fig.pde}. At this point, one can add the noise due to DCR, optical cross talk, etc. However, the noise due to DCR is sufficiently small at the LXe temperatures as to be negligible. For this studies we have not included yet the effect of optical cross-talk, which are very dependent of the specific SiPM model used for the LXSC and will, in any case, be small.  

%
%%1314_2planes_100000.root , energyDist.c
%\begin{figure}[!htb]
%	\begin{center}
%           \subfloat[XY distribution]{
%               \includegraphics[width=.5\textwidth]{img/energyXY.pdf}
%           }
%           \subfloat[XZ distribution]{
%               \includegraphics[width=.5\textwidth]{img/energyXZ.pdf}
%           }
%           \caption{\label{fig.gammas} Gamma generation from outside the box using solid angle. (a) X,Y distribution of the gammas interacting in the LXSC. (b) X,Z distribution, showing an accumulation of interactions in the first 3 cm.  }
%    \end{center}
%\end{figure}

%In order to inspect possible edge effects in the LXSC we have also run simulations changing the generation of the gammas, instead of being fired from the outside of the box within a solid angle, they start at a random point of the entry plane of the box. In this way we have events in every part of the box (see Figure \ref{fig.distNewgen}). In any case we have seen this change does not affects energy or spatial resolutions. 
%Despite of that, given the fact that algorithms to reconstruct Compton interactions are sensitive to the incoming direction we will keep generating gammas outside the box.

%lxsc2_z5.root , energyDistNewGen.c
\begin{figure}[!htb]
        \begin{center}
                \subfloat[XY distribution]{
                        \includegraphics[width=.5\textwidth]{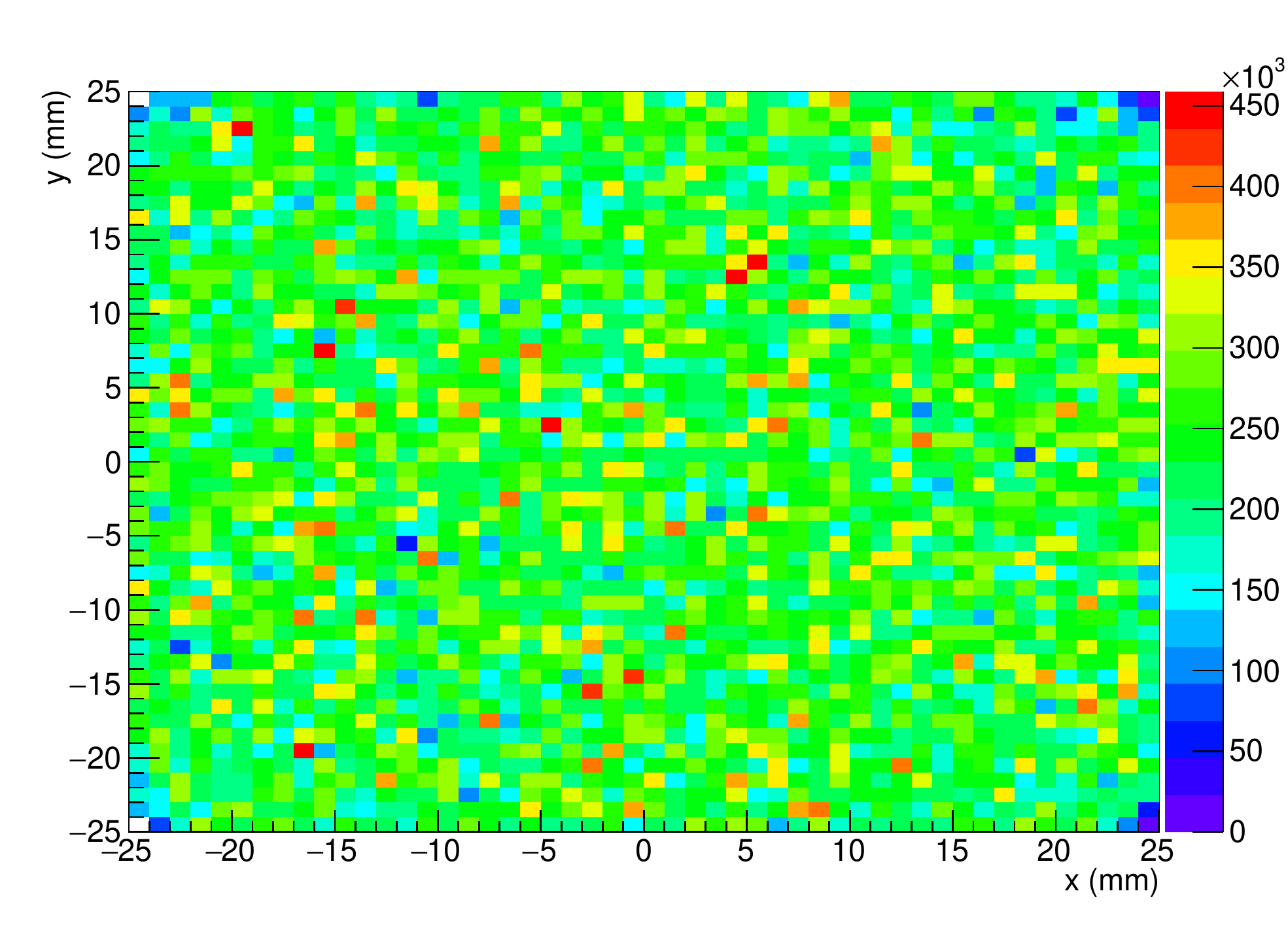}
                }
                \subfloat[XZ distribution]{
                        \includegraphics[width=.5\textwidth]{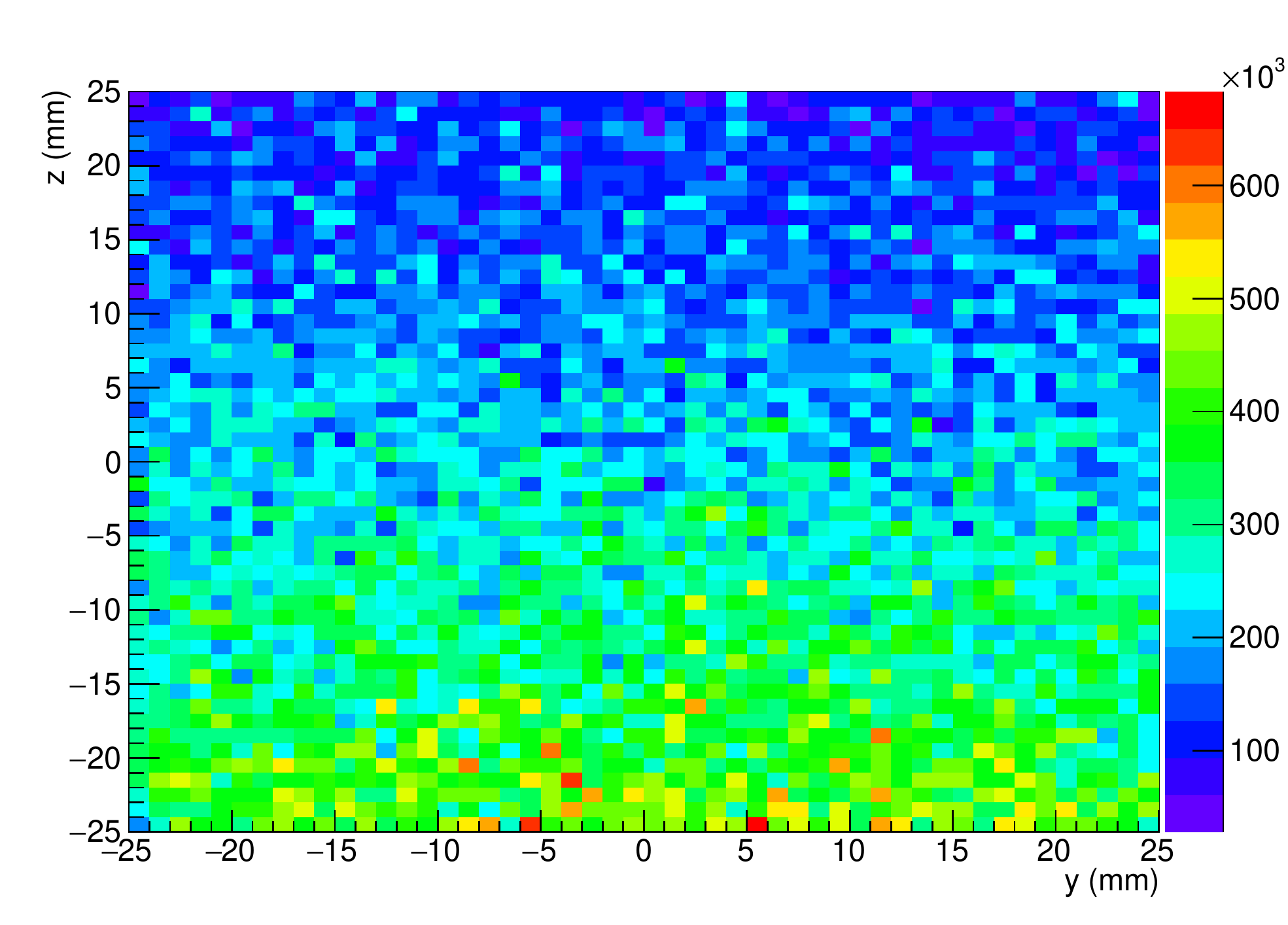}
                }
                \caption{\label{fig.distNewgen} (a) X,Y distribution of the gammas interacting in the LXSC. (b) X,Z distribution, showing an accumulation of interactions in the first 3 cm  as expected from the attenuation length.}
        \end{center}
\end{figure}

 Figure \ref{fig.distNewgen} shows the distribution of gammas generated in a specific run, distributed uniformly along x-y (transverse coordinates) and shot at $z=0$. Notice that the interactions accumulate in the firs 3 cm of the cell, as expected from the attenuation length.  
 
 \subsection{Configurations of the LXSC}

\begin{figure}[!htb]
	\centering
	\includegraphics[scale=0.5]{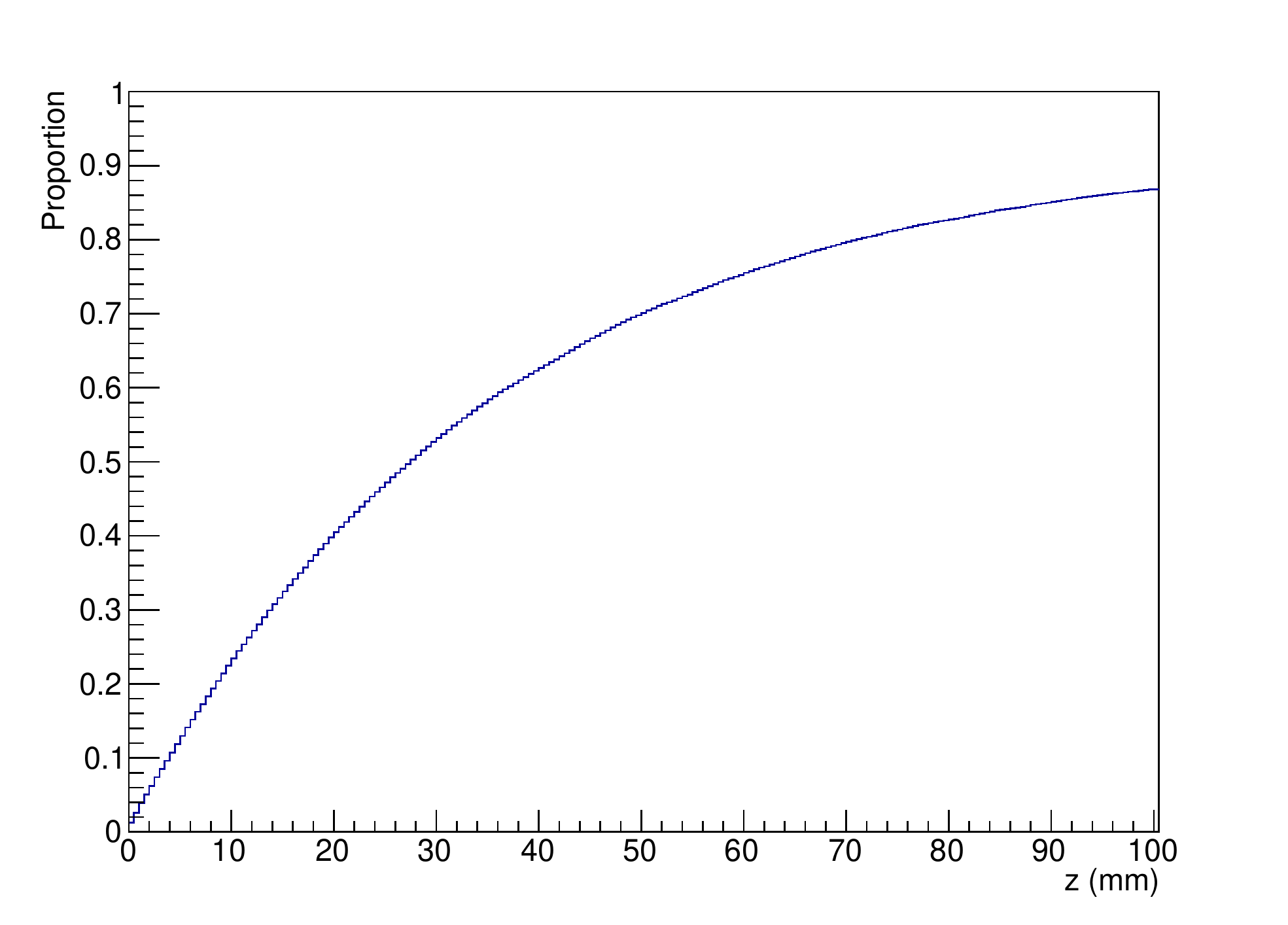}
	\caption{\label{fig.proportion} Proportion of events in function of position in $z$ coordinate. With 5 cm approximately $77\%$ of the incoming gammas interact in the active volume of the detector.}
\end{figure}

The geometry and instrumentation of the LXSC can be tuned up depending on the intended application. The most important parameters to be varied are: 
\begin{enumerate}
\item {\bf Shape and transverse dimensions}: The simplest design of the LXSC is a simple box, but other geometries, such as a trapezoid (a truncated pyramid) may be more suitable for packing the modules in a PET ring. The transverse dimensions can also be tuned up depending on the packing needs (e.g, the size of the ring). The default chosen for most of our studies for the LXSC is a box of $5 \times 5$~cm$^2$~transverse size.
\item {\bf Thickness}: our default (5 cm) represents a compromise between the conflicting requirements of good efficiency (which improves with thickness) and good spatial resolution (which worsens with thickness). With 5 cm thickness 77\% of the impending 511 keV gammas interact in the cell  (see Figure \ref{fig.proportion}).
\item {\bf Number of instrumented faces}: The best performance is obtained when all the faces are instrumented, maximizing light collection (thus energy resolution) and minimizing inhomogeneities (thus improving energy resolution and spatial resolution). On the other hand, such an arrangement is too expensive, as well as impractical for packing (e.g, arranging the boxes into a ring) purposes. Alternative configurations can instrument 4 or even 2 planes, which result, indeed, in the best overall compromise. 
\item {\bf Number of SiPMs per face}: again, the best response is achieved when the faces are fully covered by SiPMs, but in more sparse configuration reduced cost can be traded by performance.
\end{enumerate}

\subsection{Energy resolution of the LXSC}
\label{sec.energy}

%The VUV photons produced by the interactions of gammas in the LXSC are shifted to 420 nm by the TPB which coats the teflon panels and the SiPMs. The conversion efficiency, measured, among other authors by the NEXT collaboration is 80\%. The resulting blue photons are registered by the SiPMs, with high efficiency (the typical PDE of modern SiPMs for 420 nm exceeds 50\%. For this specific study we have simulated the 6 mm SENSL C-series sensor). 
%
%While the fast decay time of scintillation (2 ns) allows for TOF applications (see Section \ref{sec.tof}), the longer time associated with the recombination of electrons (45 ns) implies integration times of the order of 200-300 ns, similar to those used by LSO detectors. The dark current of the SiPMs and the electronic noise of the front-end electronics (in the case than an ASIC is used) is very low due to the cryogenic operation, and thus the resolution is dominated by photoelectron statistics. 

\begin{figure}[!htb]
	\centering
	\subfloat[LXSC6-64]{
		\includegraphics[width=.55\textwidth]{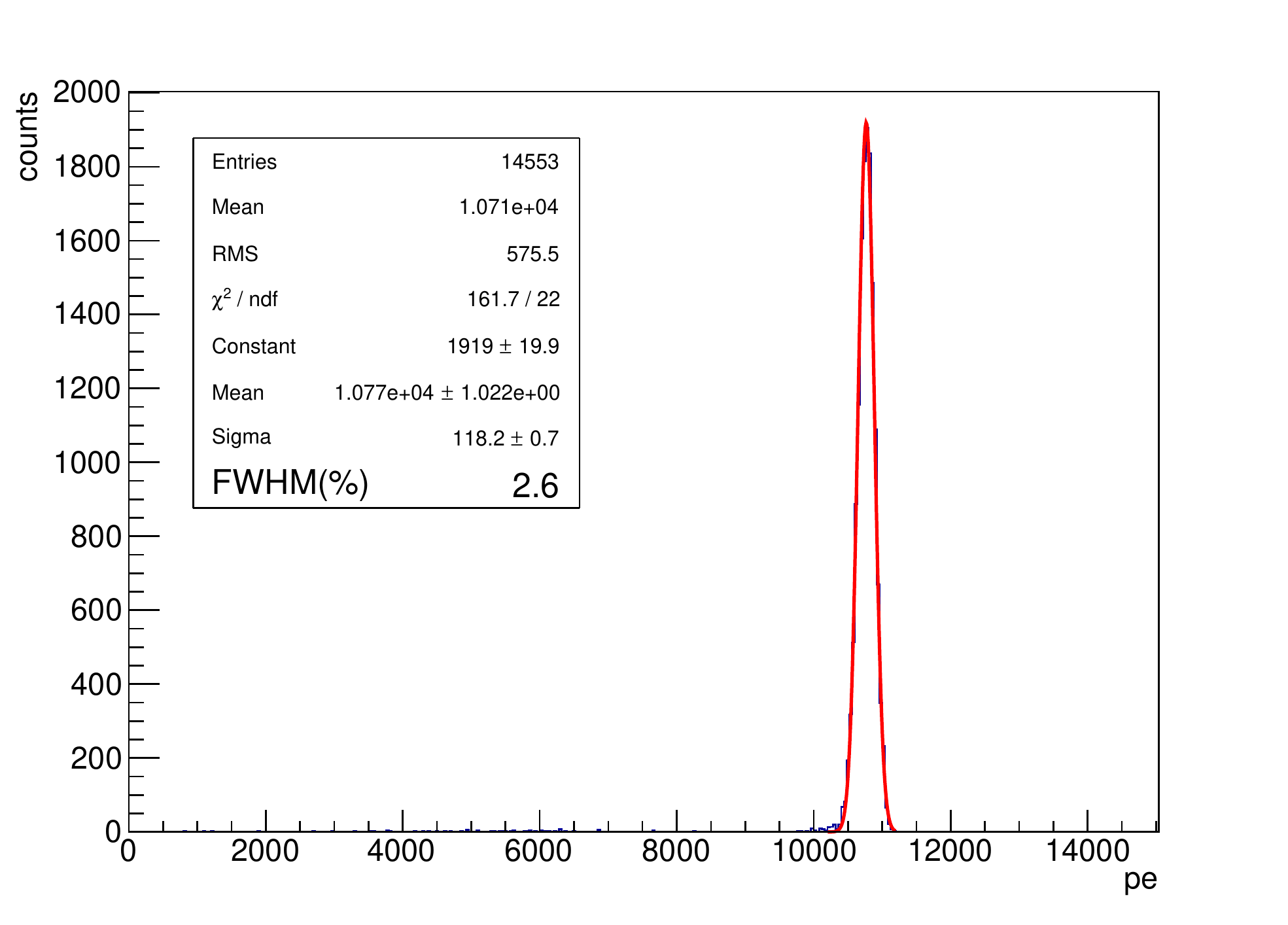}
		\label{fig.energy664}
	}
	\subfloat[LXSC2-36]{
		\includegraphics[width=.55\textwidth]{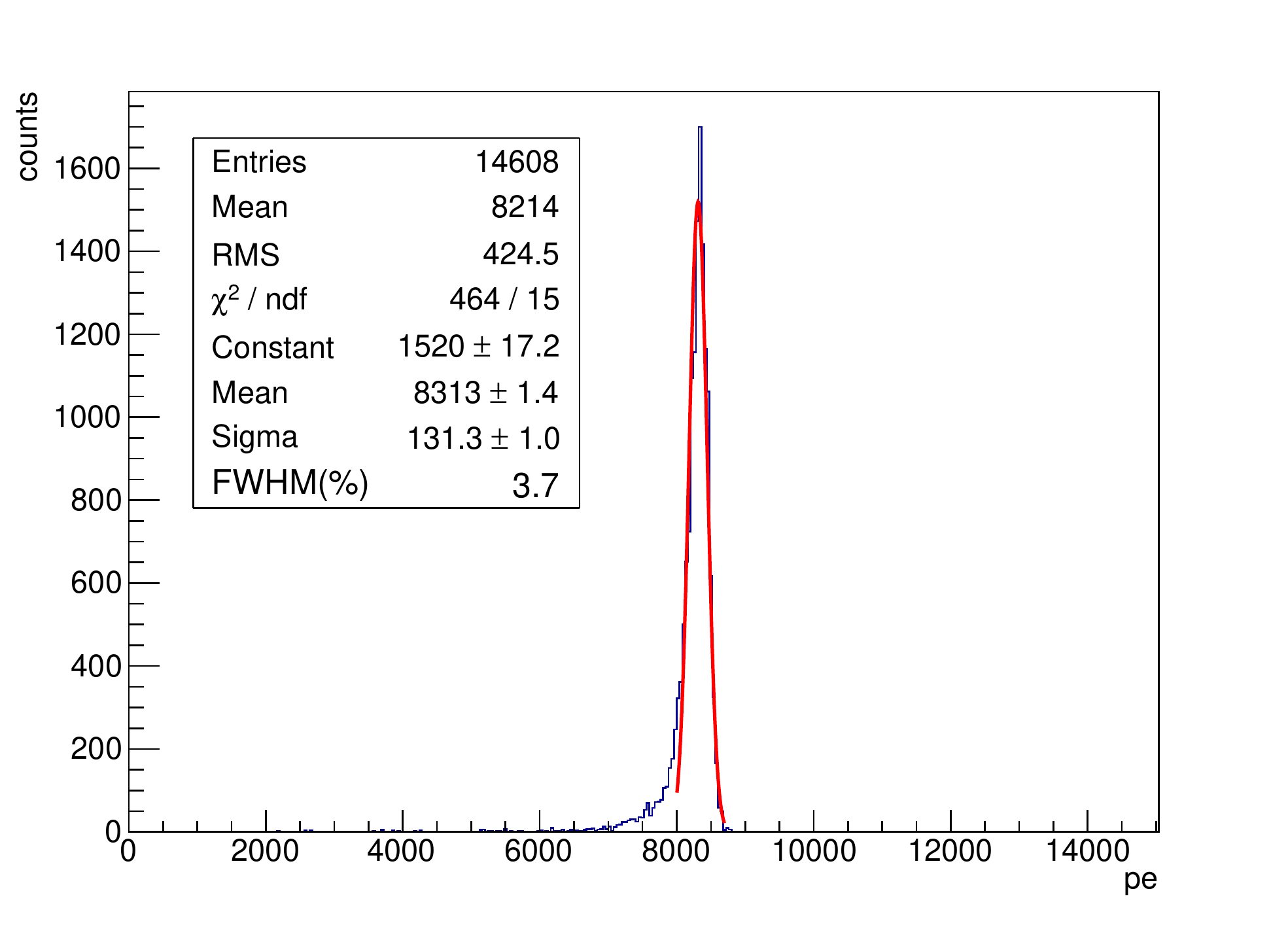}
		\label{fig.energy236}
	}
	\caption{\label{fig.energy} (a) The intrinsic energy resolution of the LXSC6 is excellent (2.6\% FWHM) due to the high photoelectron statistics. (b) The most sparse configuration, LXSC2 records less light, but the resolution is still very good (3.7\% FWHM).}
\end{figure}

Figure \ref{fig.energy664} shows the recorded number of photoelectrons in the LXSC6-64 (6 instrumented faces with 64 SiPM each of them) corresponding to photoelectric interactions. The fit shows a resolution of 2.6\% FWHM, {\em much better} than the resolution obtained with conventional SSDs (for example, the best resolution obtained with test systems for LSO crystals is in the vicinity of 9 \% FWHM, while commercial PETs typically show a resolution of around 20 \% FWHM). Figure \ref{fig.energy236} shows the recorded number of photoelectrons in the LXSC2-36 (2 instrumented faces with 36 SiPM each of them) corresponding to photoelectric interactions ($\sim$ 8000 P.E., to be compared with $\sim$ 11000 P.E. for the LXSC6). The  fit yields a resolution of 3.7\% FWHM. This is still very good, but it can be further improved by applying geometrical corrections.

\begin{figure}[!htb]
	\centering
			\subfloat[LXSC6, $x$ coordinate]{
				\includegraphics[width=.5\textwidth]{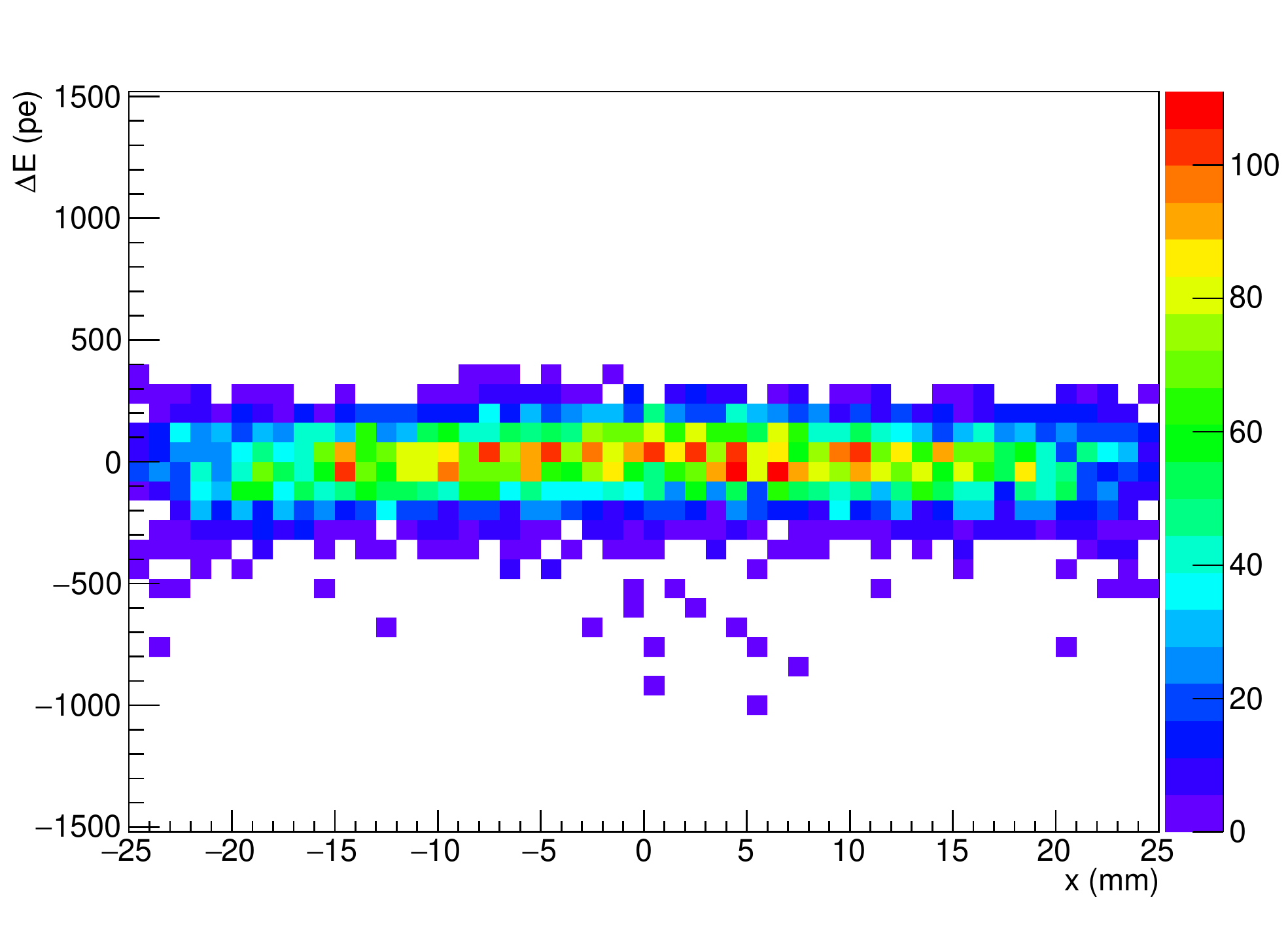}
				\label{fig.eposx664}
			}
			\subfloat[LXSC6, $z$ coordinate]{
				\includegraphics[width=.5\textwidth]{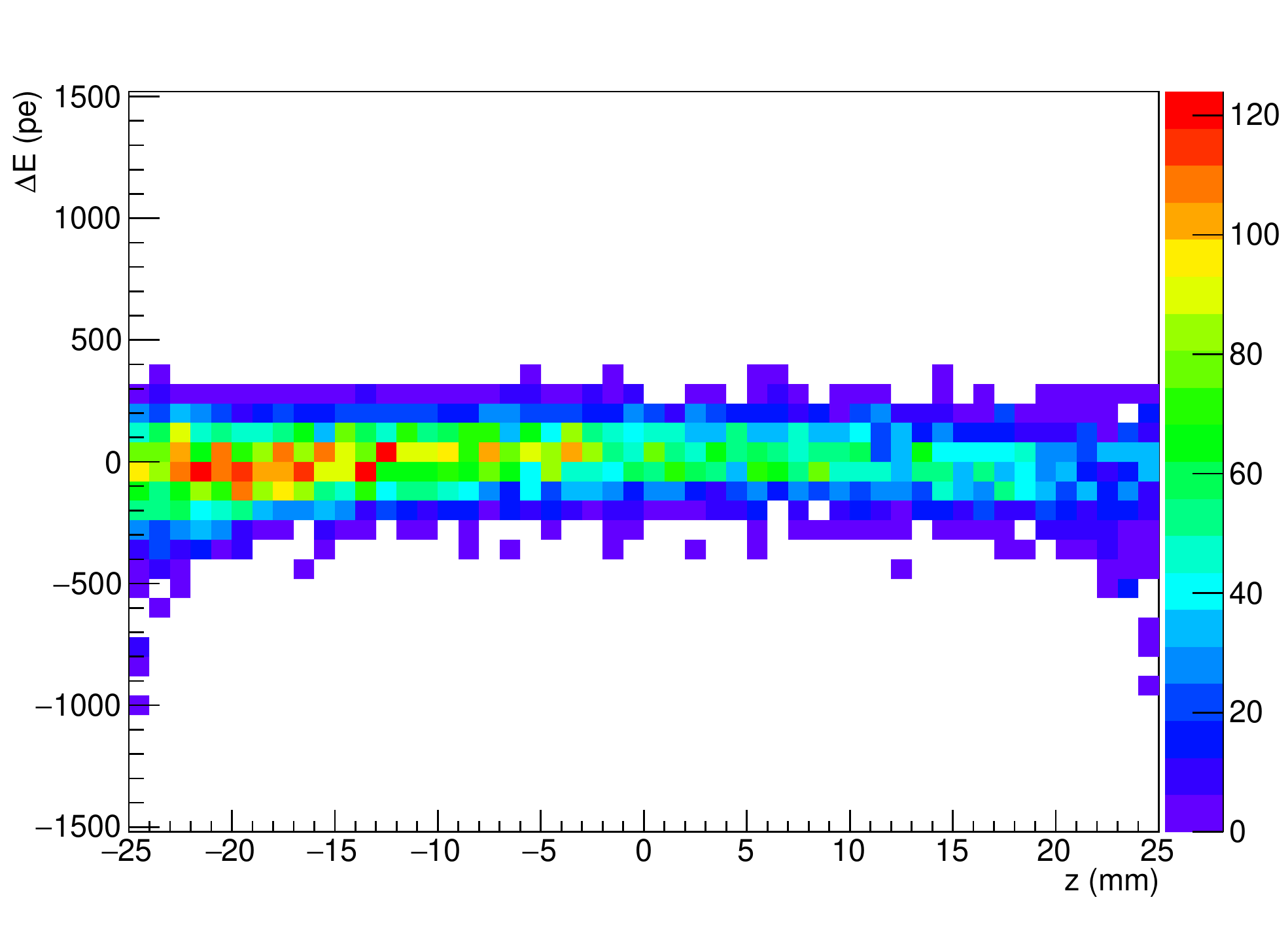}
				\label{fig.eposz664}				
			}\\
	\caption{\label{fig.energyDep6} Difference between the reconstructed energy and the average energy ($\Delta E$) as a function of one coordinate. (a) Energy dependence with the transverse coordinate ($x$) in LXSC6. (b) Energy dependence with the longitudinal coordinate ($z$) in LXSC6.}
\end{figure}

\begin{figure}[!htb]
	\centering			
		\subfloat[LXSC4, $x$ coordinate]{
			\includegraphics[width=.5\textwidth]{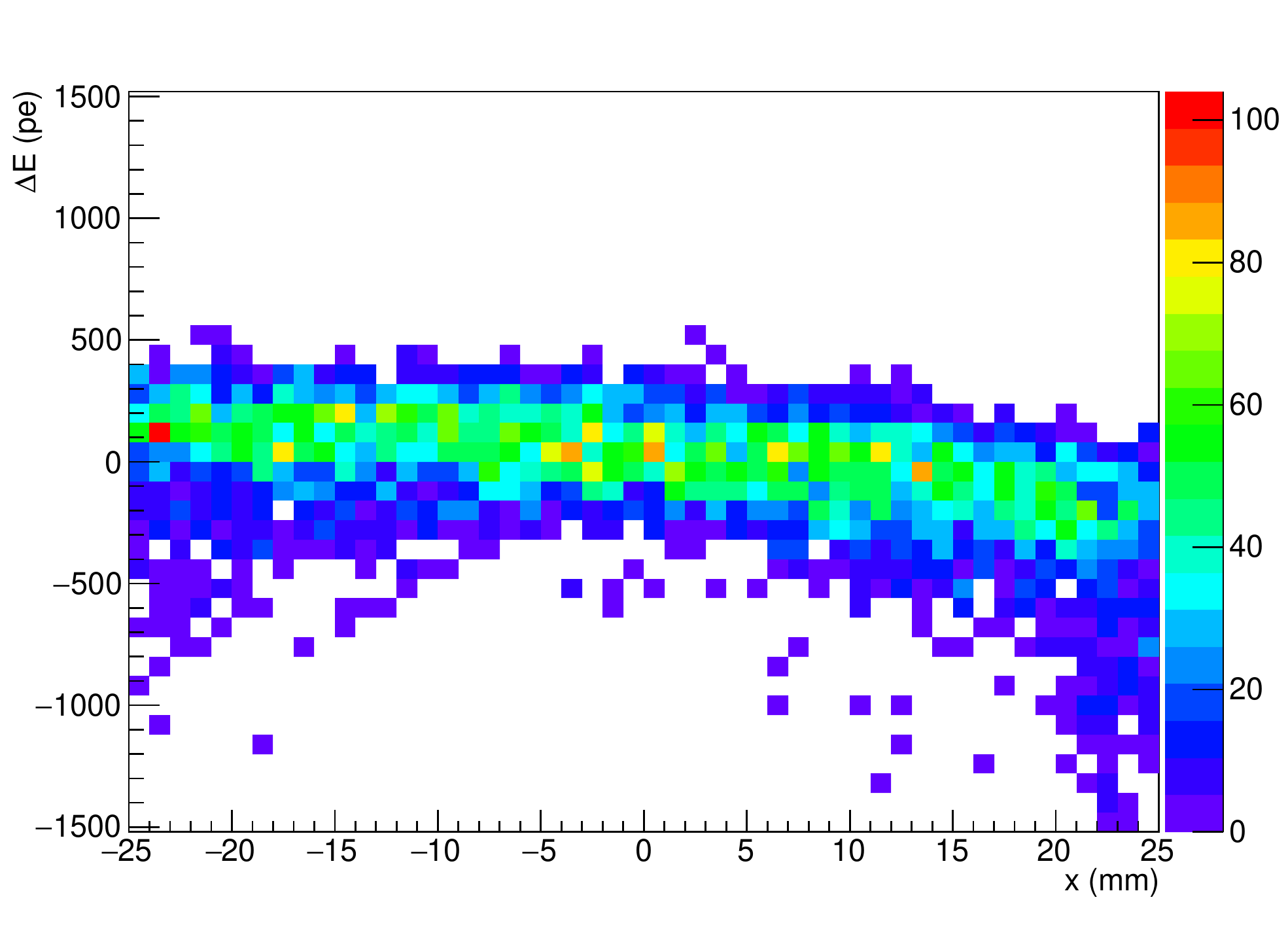}
			\label{fig.eposx464}
		}
		\subfloat[LXSC4, $z$ coordinate]{
			\includegraphics[width=.5\textwidth]{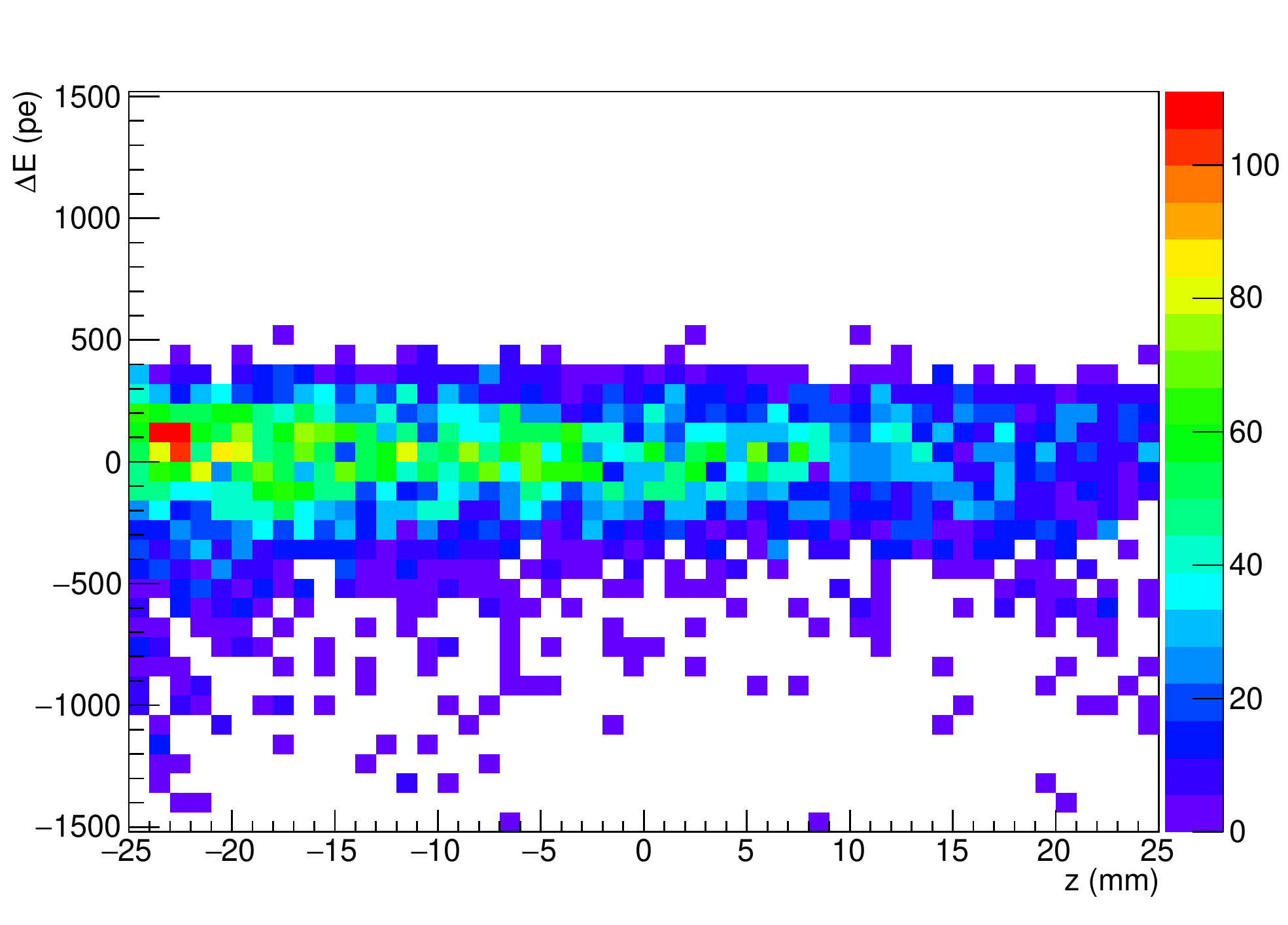}
			\label{fig.eposz464}				
		}\\
	\caption{\label{fig.energyDep4} Difference between the reconstructed energy and the average energy ($\Delta E$) as a function of one coordinate. (a) Energy dependence with the transverse coordinate ($x$) in LXSC4. (b) Energy dependence with the longitudinal coordinate ($z$) in LXSC4.}
\end{figure}

\begin{figure}[!htb]
	\centering		
		\subfloat[LXSC2, $x$ coordinate]{
			\includegraphics[width=.5\textwidth]{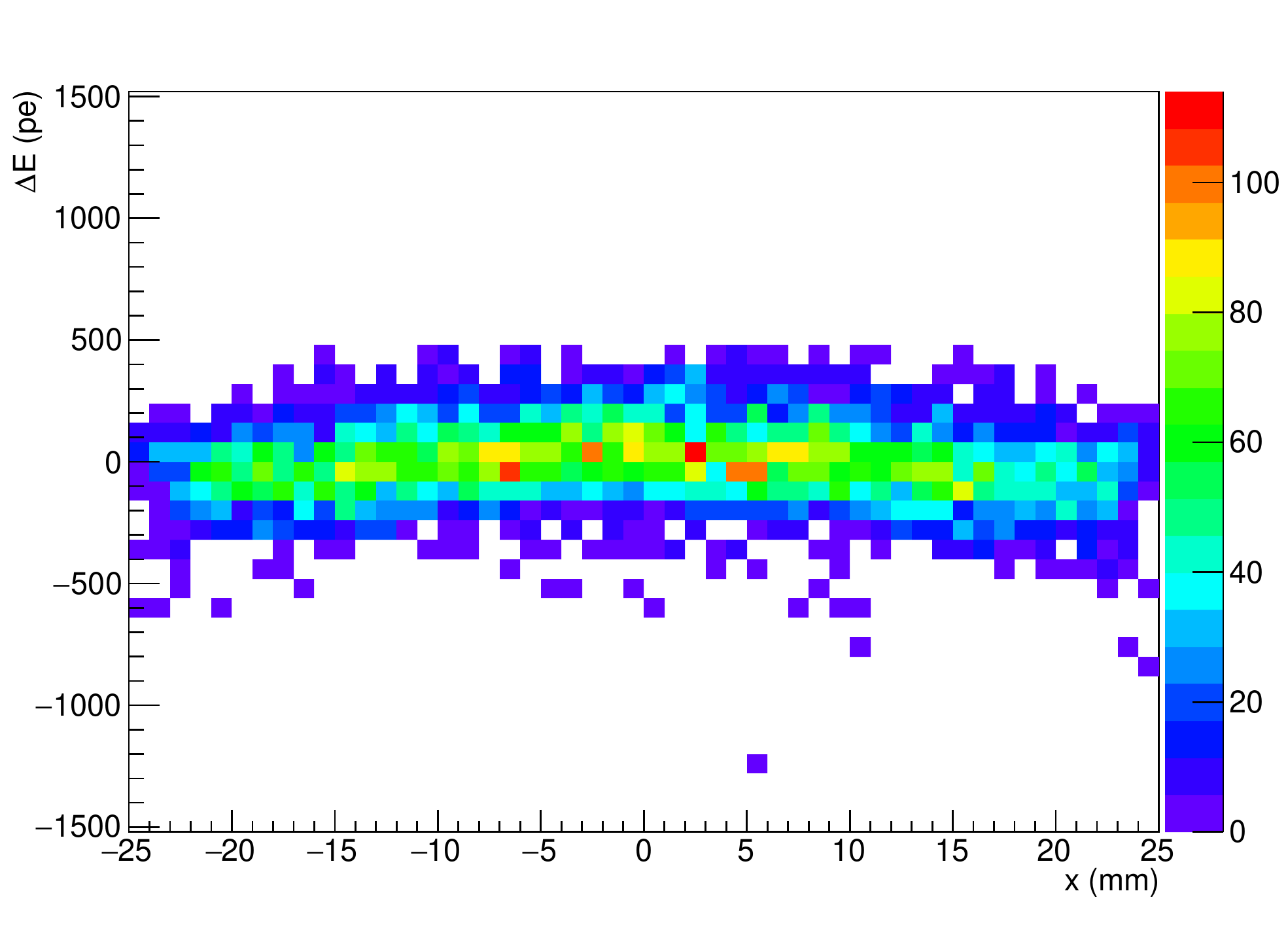}
			\label{fig.eposx264}
		}
		\subfloat[LXSC2, $z$ coordinate]{
			\includegraphics[width=.5\textwidth]{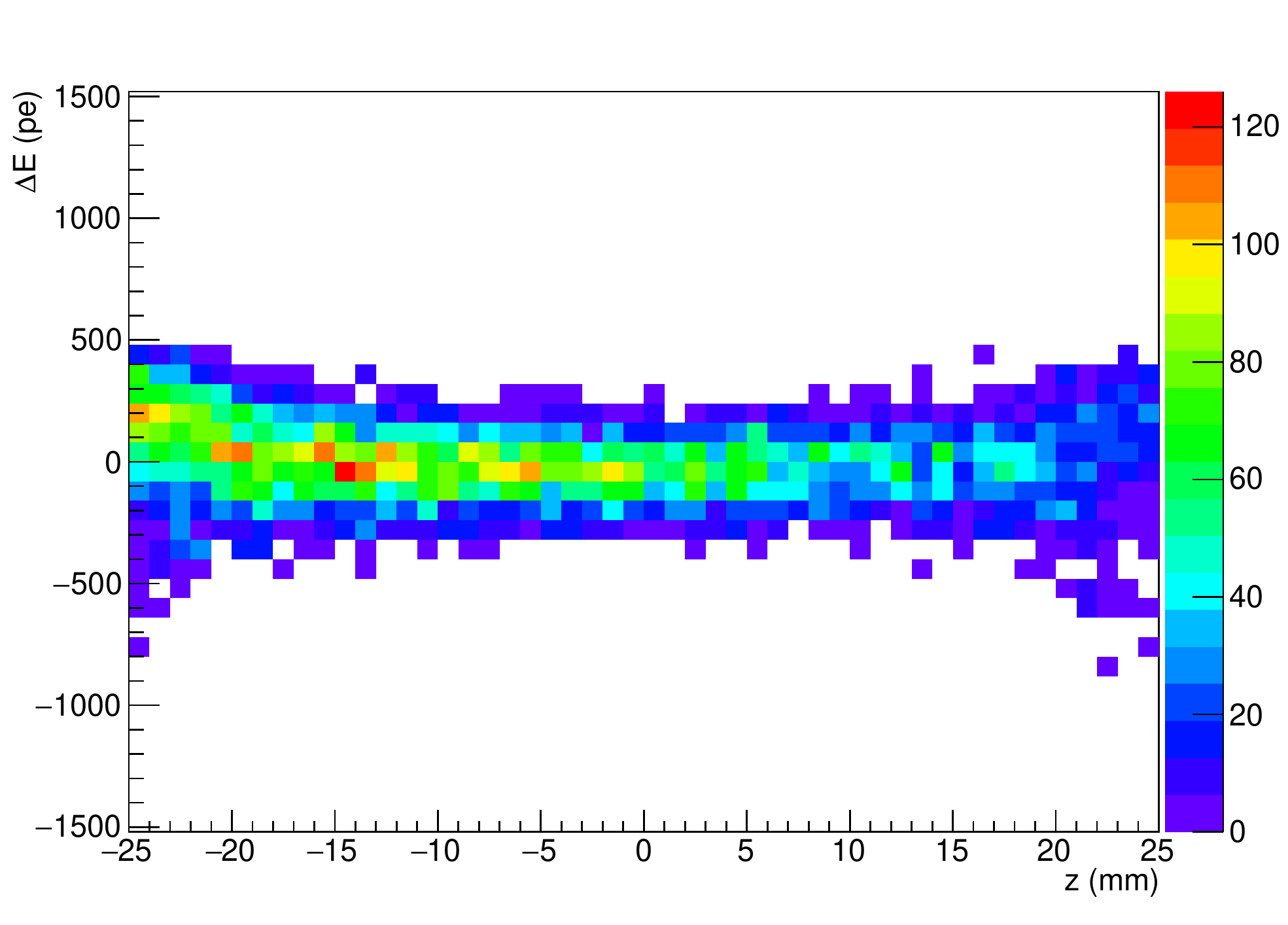}
			\label{fig.eposz264}
		}
	\caption{\label{fig.energyDep2} Difference between the reconstructed energy and the average energy ($\Delta E$) as a function of one coordinate. (a) Energy dependence with the transverse coordinate ($x$) in LXSC2. (b) Energy dependence with the longitudinal coordinate ($z$) in LXSC2.}
\end{figure}

The need for geometrical corrections is illustrated in Figures \ref{fig.energyDep6},\ref{fig.energyDep4} and \ref{fig.energyDep2}, which shows the difference between reconstructed energy and average energy as a function of one coordinate. The dependence of the energy with the longitudinal and transverse coordinate in the LXSC6 is very soft (Figures \ref{fig.eposx664} and \ref{fig.eposz664}), and can be neglected for all practical purposes. This is indeed expected, as the LXSC6 is totally symmetric and solid angle effects are minimized. If we decrease the amount of instrumentation, energy resolution worsens to around $\sim4\%$ for 49 SiPM and $\sim5\%$ for 36 SiPM. This is due to the geometrical corrections needed as we have covered less surface with sensors.

In the case of the LXSC4 (Figures \ref{fig.eposx464} and \ref{fig.eposz464}) the effect is larger due to the asymmetry of the detector giving resolutions of $\sim 4\%$. LXSC2 requires smaller corrections as is shown in Figures \ref{fig.eposx264} and \ref{fig.eposz264}, resolution in this case is $\sim 3.7\%$. Applying geometrical corrections one could improve the resolutions quoted.

%\begin{table}[h]
%\caption{\label{tab.energy1} Energy resolution (\% FWHM) for LXSC in function of the number of planes and the number of SiPM on each plane.}
%\begin{center}
% \begin{tabular}{c|ccc}
%  \toprule
%   Planes\textbackslash SiPM & \textbf{36 SiPM} & \textbf{49 SiPM} & \textbf{64 SiPM} \\
%   \hline
%  {\bf 2 Planes} & 3.7\% & 3.5\% & 3.7\%\\
%  {\bf 4 Planes} & 4.4\% & 4.1\%  & 3.9\%\\
%  {\bf 6 Planes} & 5.1\% & 4.1\% & 2.6\%\\
%  \toprule
% \end{tabular}
%\end{center}
%\end{table}

%\begin{table}[h]
%\begin{center}
%\caption{\label{tab.energy2} Energy resolution for LXSC2 varying longitudinal size.}
%  \begin{tabular}{c|c}
%   \toprule
%    Longitudinal size & \textbf{Resolution (FWHM)} \\
%    \hline
%   {\bf 2 cm} & 5.0\% \\
%   {\bf 3 cm} & 4.5\% \\
%   {\bf 4 cm} & 3.9\% \\
%   {\bf 5 cm} & 3.8\% \\
%   \toprule
%  \end{tabular}
% \end{center}
%\end{table}

Reducing the longitudinal size of the box has also an effect on resolution. We have studied this effect on LXSC2. Resolution is 3.7\% for 5 cm and 5\% for 2 cm. The resolution worsens due to energy losses, but is still acceptable. 

Finally, if we assume the lowest $W_{ph}$~measured for electrons
(24,000 photons rather than 37,000 photons for a 511 keV gamma)
 the yield would be reduced by 65\% and the resolution of the LXSC2 would be spoiled by a factor $1./\sqrt{0.65} = 1.2$. One would then have a resolution of around 4.5\% FWHM, still much better than that of modern SSDs such as LSO/LYSO.

\begin{figure}[!htb]
	\centering
	\subfloat[LXSC6-64]{
		\includegraphics[width=.33\textwidth]{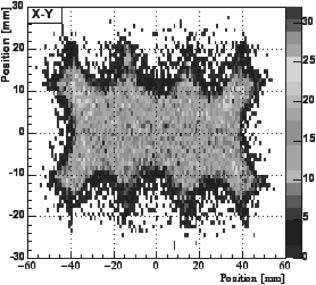}
	}
	\subfloat[LXSC2-36]{
		\includegraphics[width=.33\textwidth]{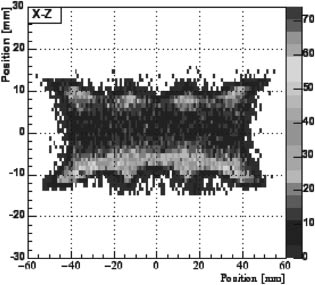}
	}
	\subfloat[LXSC2-36]{
		\includegraphics[width=.33\textwidth]{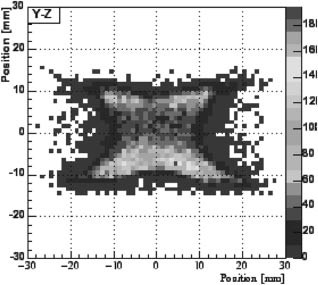}
	}
	\caption{\label{fig.wasedaGeo} Position distributions of interaction points for annihilation gamma rays in the Waseda cell \cite{nishikido05}. Very large geometrical corrections are needed. (a) XY distribution, (b) XZ distribution and (c) YZ YZ distribution.}
\end{figure}

Notice that the small geometrical corrections found in the LXSC are a crucial difference with the Waseda cell, where the geometrical corrections, due to the large size of the PMTs where very large (see Figure \ref{fig.wasedaGeo}) and made, ultimately, the cell impractical as a detection device, since only the central part of the detector ($5 \times 5 \times 5$~mm$^3$) was useful. The second crucial difference is that the LXSC registers much more light than the Waseda cell. This is due to the fact that all the faces are reflecting (the Waseda cell left one face open, resulting in large losses and fluctuations) and to the use of SiPMs, which have very large PDE ($\sim$ 50\% to be compared with the 5-20\% of the Waseda PMTs) right in the region (420 nm) where the light is shifted from TPB. 

%\newpage\null\thispagestyle{empty}\newpage
%\thispagestyle{newstyle}
\subsection{Spatial resolution of the LXSC}
\label{sec.spatial}

\subsubsection*{Barycenter algorithm}
The simplest way to determine the point of interaction of an incoming photon, $(x,y,z)$~ is to use a barycenter algorithm. 
\[
\xi_r = \frac{\sum \xi_i N_i}{N}
\]
where $\xi_r$~stands for each one of the three coordinates ($x_r, y_r, z_r$), $N_i$~is the number of photoelectrons registered in each SiPM and $N=\sum N_i$. In the LXSC6 and LXSC4 configurations one can compute redundant measurements of $\xi_r$~for each coordinate. In the LXSC2 one can obtain ($x_r,y_r$) redundantly from the information found in the entry and exit faces and $z_r$~from the ratio of energy measured in the entry and the exit faces. To get a bound for the spatial resolution in our simulations we have chosen the best value for each coordinate. This could also be approximated using the charge recorded by each plane as a estimator of distance, thus we can select the best plane for each case as the one with more charge recorded.

\begin{figure}[H]
	\centering
	\subfloat[][Charge recorded in entry plane ($z=-25$) \\ for an event at $z=-18.81$.]{
		\includegraphics[width=.5\textwidth]{img/photoelectric_in_znear.pdf}
		\label{fig.photoelectricA}
	}
	\subfloat[][Charge recorded in exit plane ($z=25$) \\ for an event at $z=-18.81$.]{
		\includegraphics[width=.5\textwidth]{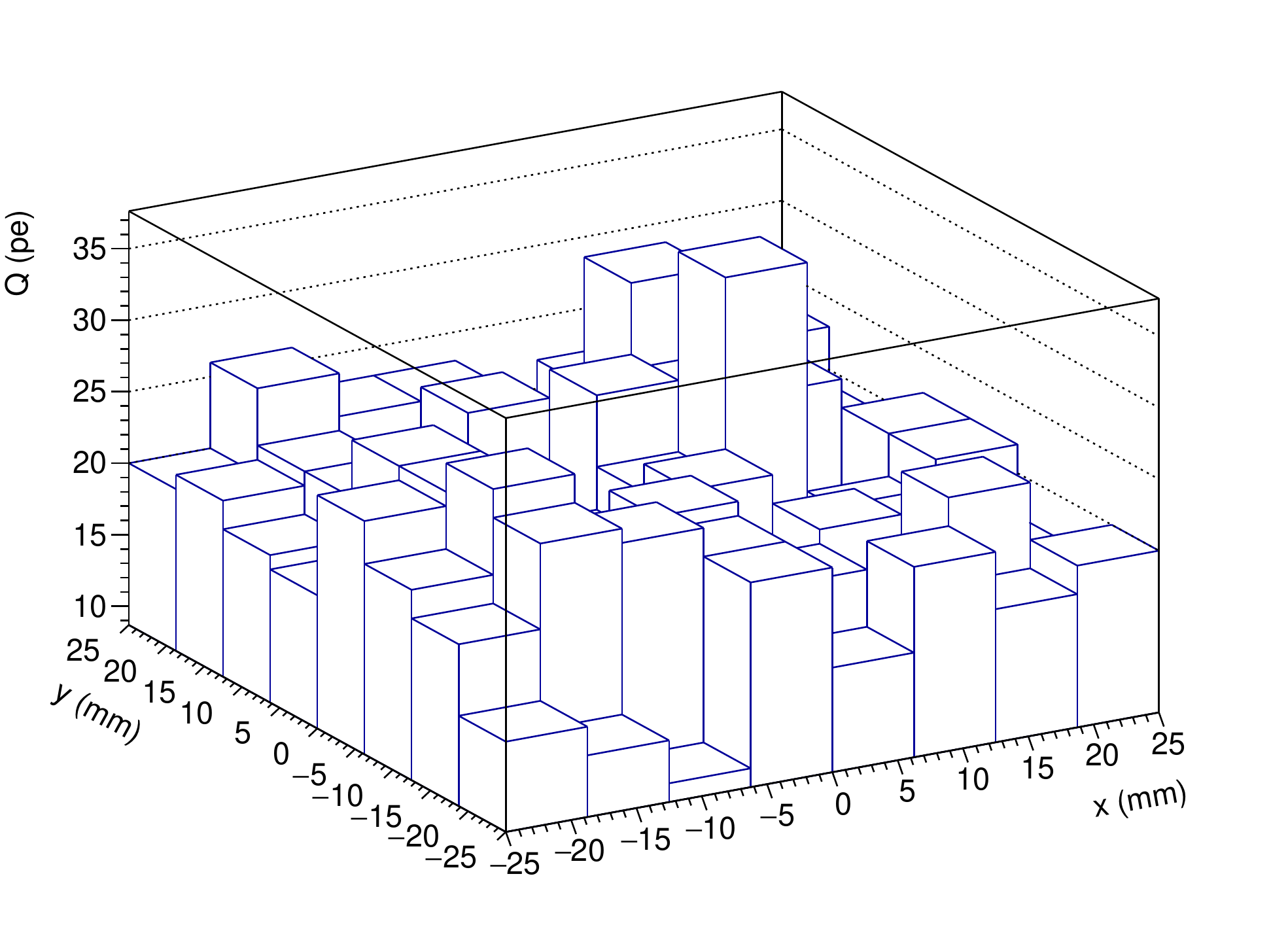}
		\label{fig.photoelectricB}
	}\\
	\vspace{-0.5cm}
	\subfloat[][Charge recorded in entry plane ($z=-25$) \\ for an event at $z=1.96$.]{
		\includegraphics[width=.5\textwidth]{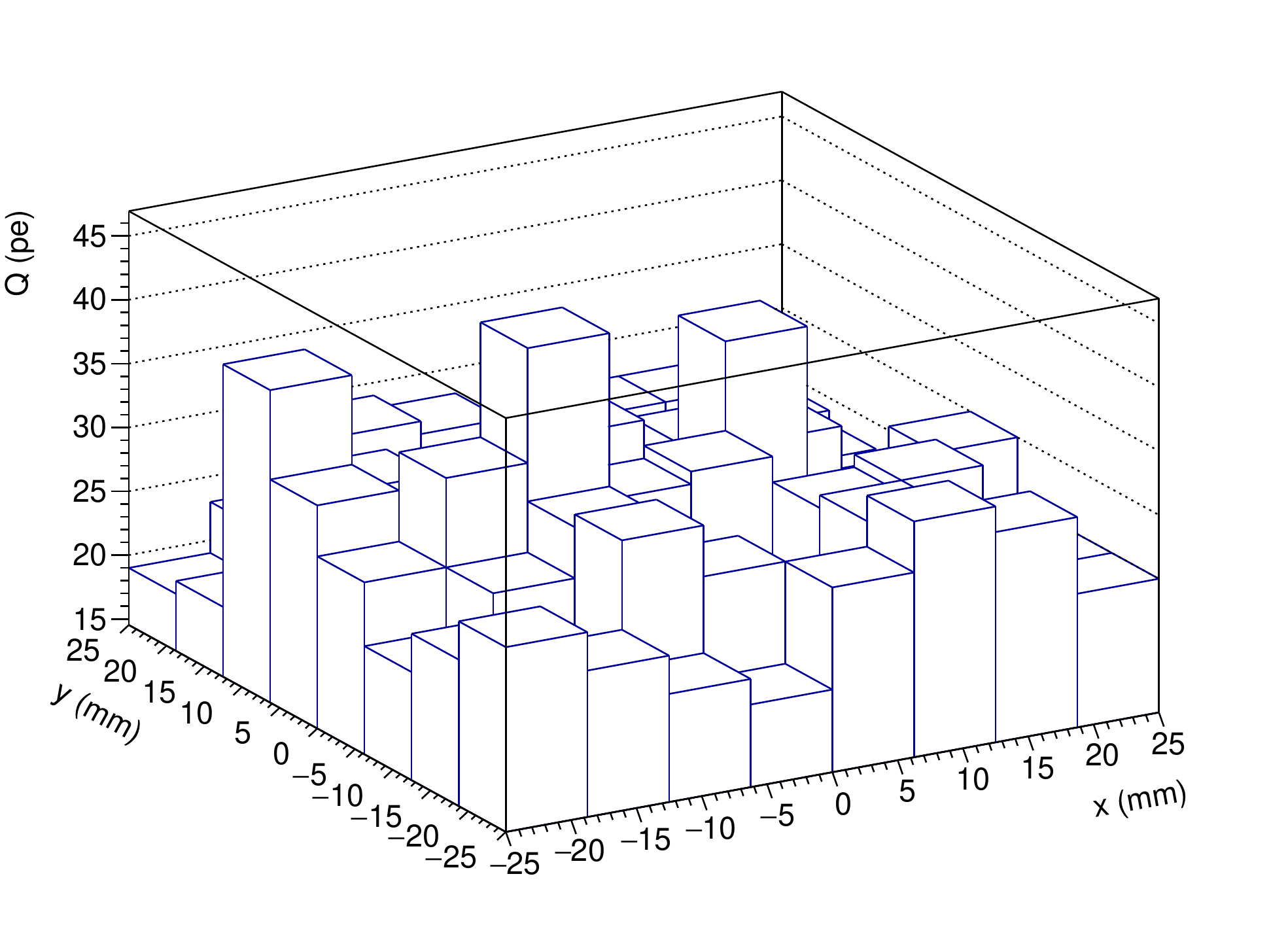}
		\label{fig.photoelectricC}
	}
	\subfloat[][Charge recorded in exit plane ($z=25$) \\ for an event at $z=1.96$.]{
		\includegraphics[width=.5\textwidth]{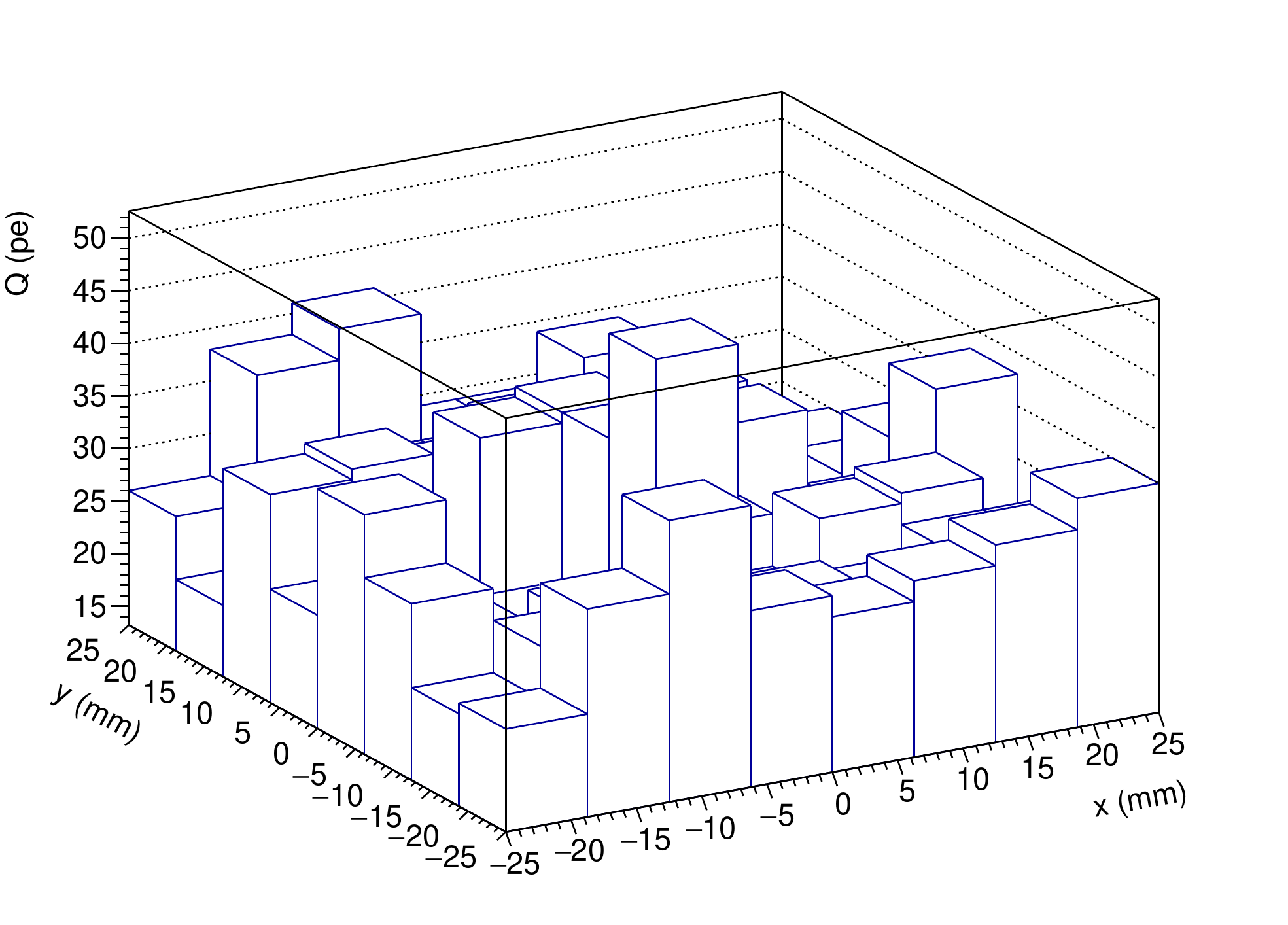}
		\label{fig.photoelectricD}
	}\\
	\vspace{-0.5cm}
	\subfloat[][Charge recorded in entry plane ($z=-25$) \\ for an event at $z=16.89$.]{
		\includegraphics[width=.5\textwidth]{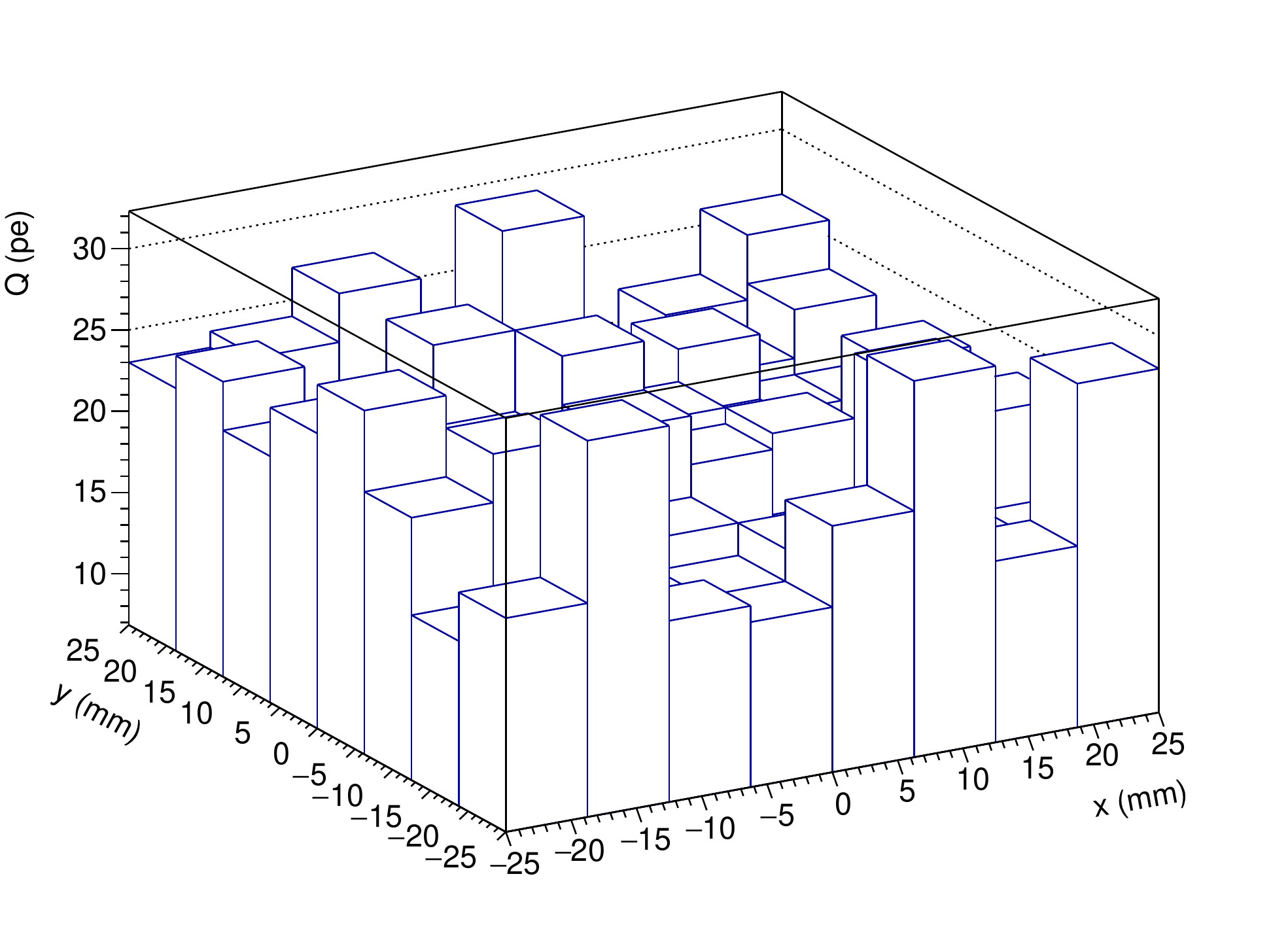}
		\label{fig.photoelectricE}
	}
	\subfloat[][Charge recorded in exit plane ($z=25$) \\ for an event at $z=16.89$.]{
		\includegraphics[width=.5\textwidth]{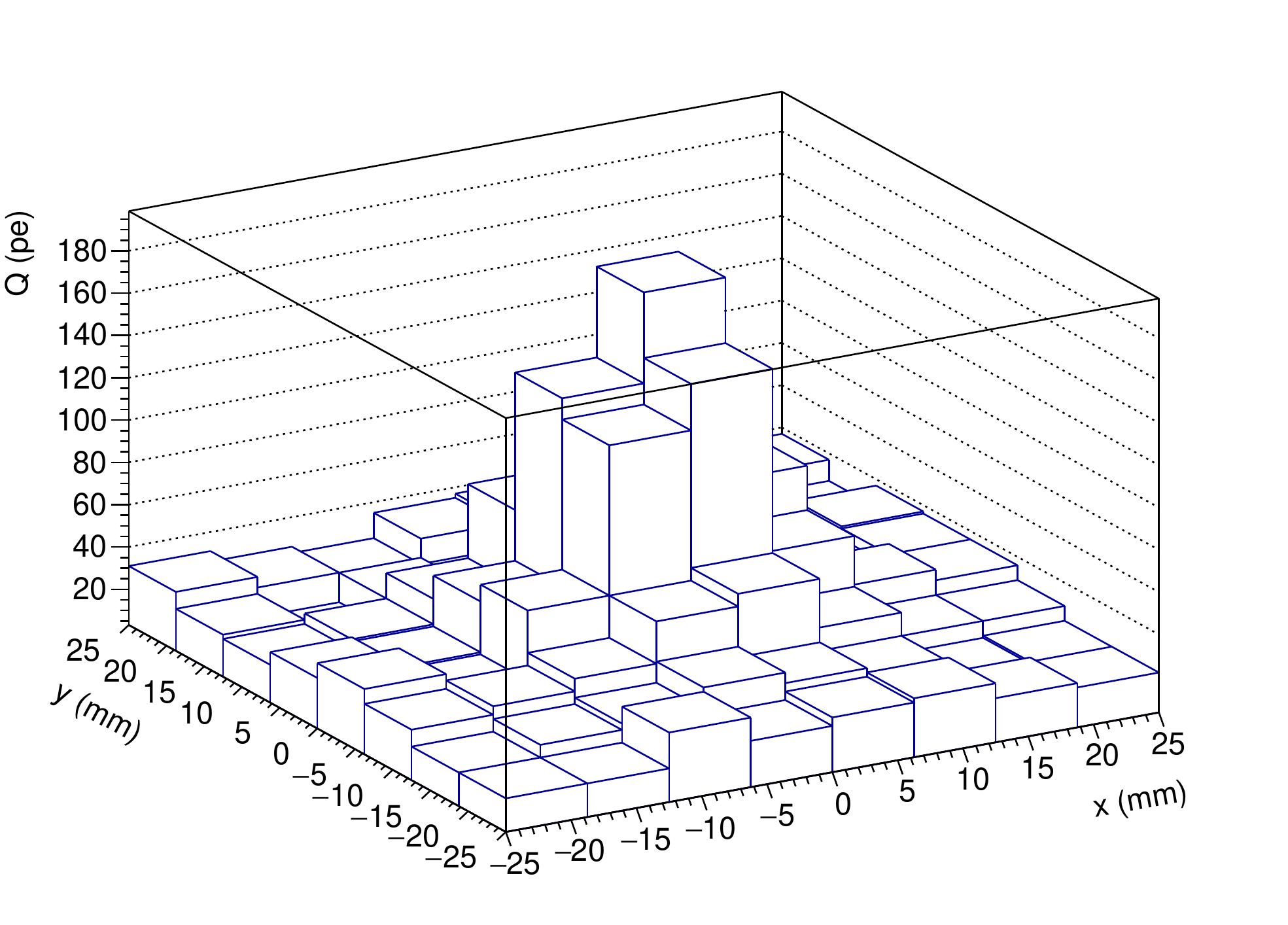}
		\label{fig.photoelectricF}
	}\\
	\caption{ \label{fig.photoelectric} Charge deposited on entry/exit plane for photoelectric events at several positions. Each row shows the same event as it is seen on each plane (entry on the left and exit on the right). First row shows an event at (1.26, 2.62,-18.81), second row at (0.63,-1.36,1.96) and the third at (0.88,0.46,16.89).}
\end{figure}

Figures \ref{fig.photoelectricA} and \ref{fig.photoelectricB} show the signal recorded by entry and exit planes respectively for one event a few millimiters away from entry plane. The interaction vertex is clearly visible in the entry plane but not in the other. If the interaction takes place near the center of the box, then the vertex is blurred in both planes (Figures \ref{fig.photoelectricC} and \ref{fig.photoelectricD}). Finally, the effect is the opposite for an event taking place near the exit plane (Figures \ref{fig.photoelectricE} and \ref{fig.photoelectricF}). 

\begin{figure}[!htb]
	\centering
	\subfloat[Entry plane]{
		\includegraphics[width=.5\textwidth]{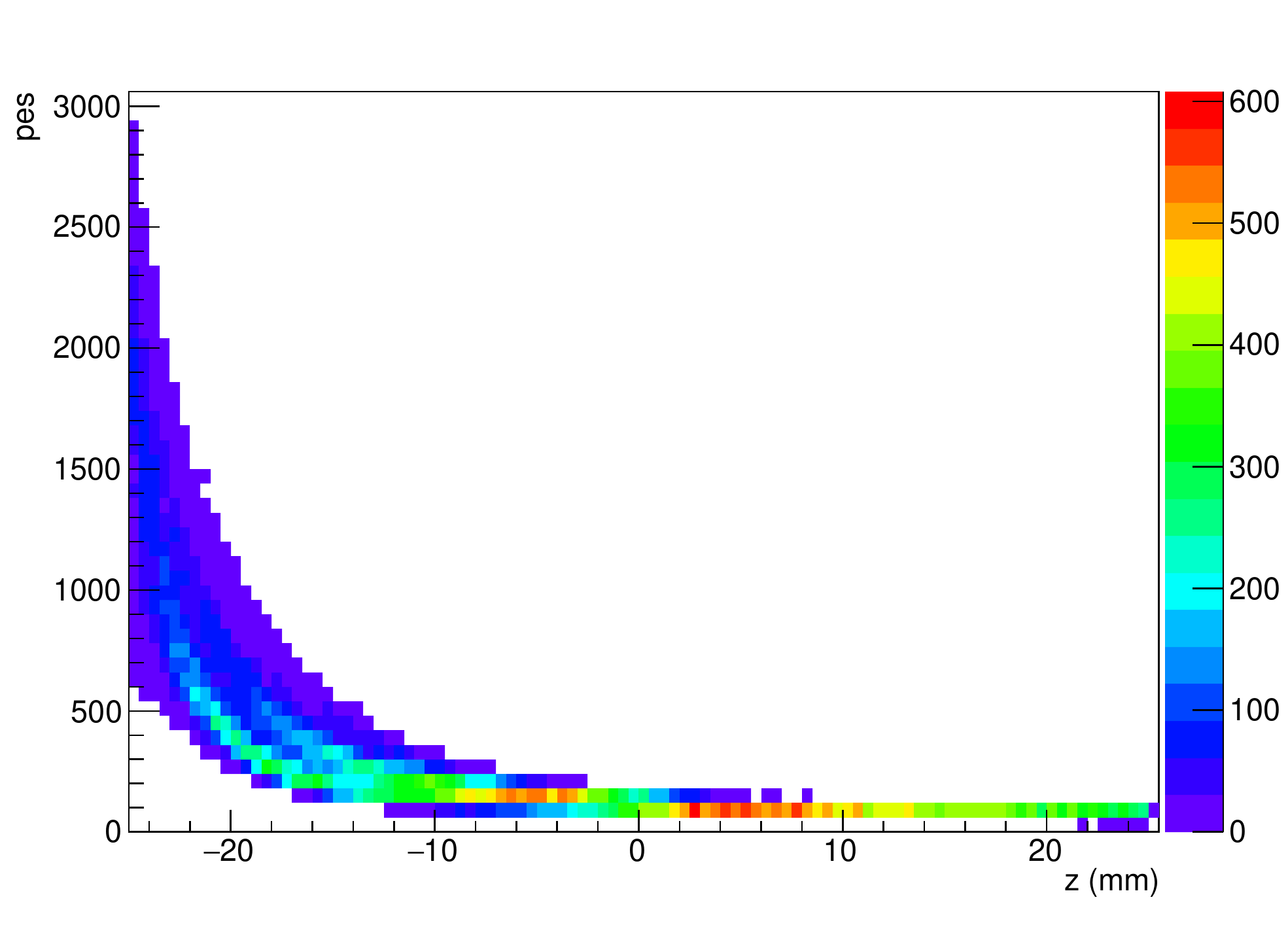}
		\label{fig.simpmmc_p0}
	}
	\subfloat[Exit plane]{
		\includegraphics[width=.5\textwidth]{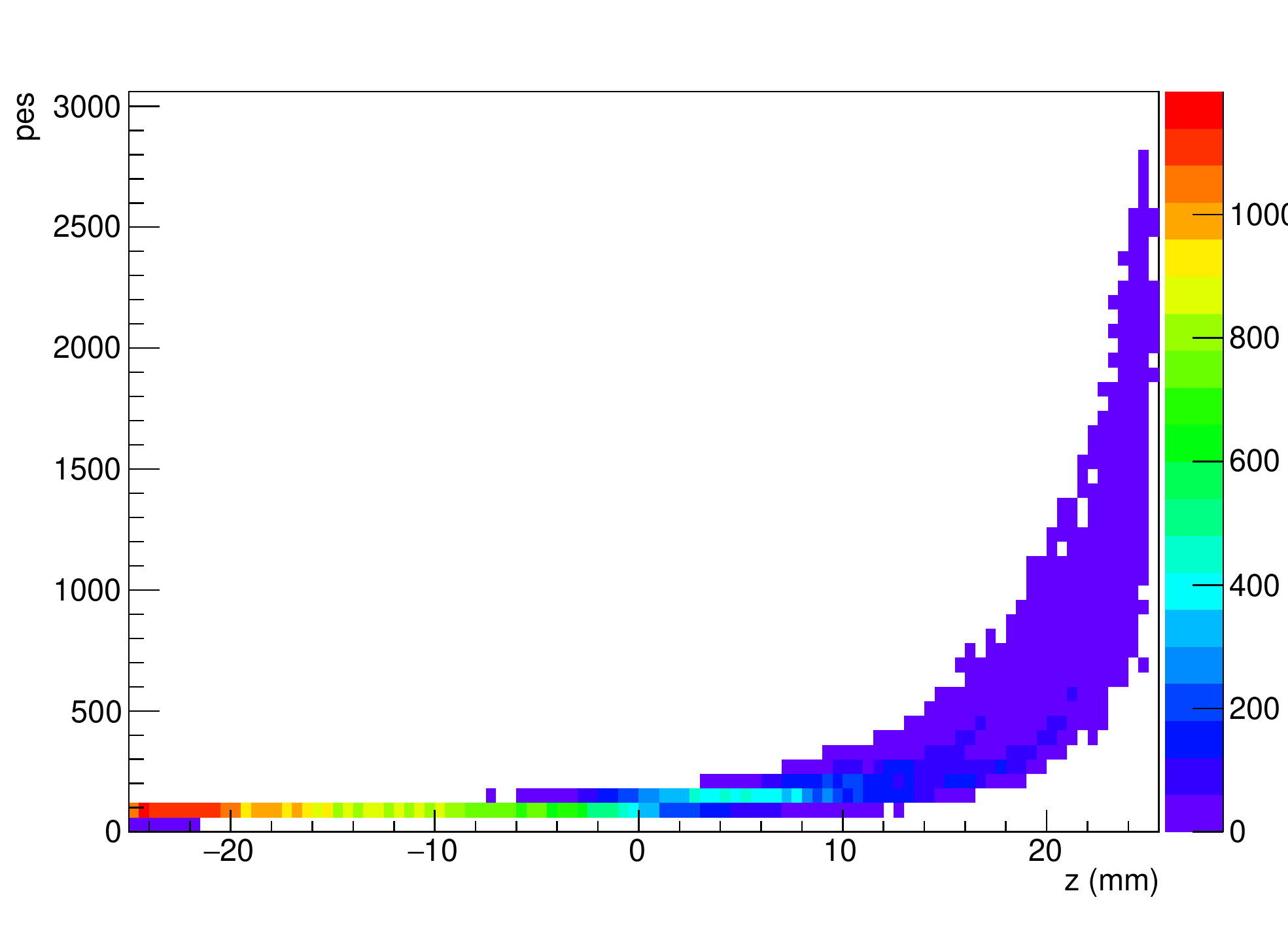}
		\label{fig.simpmmc_p2}
	}
	\caption{\label{fig.sipmm} (a) Number of photoelectrons in the SiPM registering the maximum signal as function of the distance to the entry plane ($z=-25$). (b) Number of photoelectrons in the SiPM registering the maximum signal as a function of the distance to the exit plane ($z=25$).}
\end{figure}

The effect of the box thickness is illustrated quantitatively in Figure \ref{fig.sipmm}, which shows the number of photoelectrons in the SiPM registering the maximum signal (max signal sensor or MSS) as a function of the distance to the entry face (\ref{fig.simpmmc_p0}) and the number of photoelectrons in the SiPMs registering the maximum signal as a function of the distance to the exit face (\ref{fig.simpmmc_p2}). The signal of the MSS stays above 100 pes for the first (and the last) 2 cm of the cell, therefore we can use the entry face signal in one case, and the exit face signal in the other. In the central volume of about 1 cm, the position can be determined by the combination of the signal found in both faces.

\begin{figure}[!htb]
	\centering
	\includegraphics[scale=0.5]{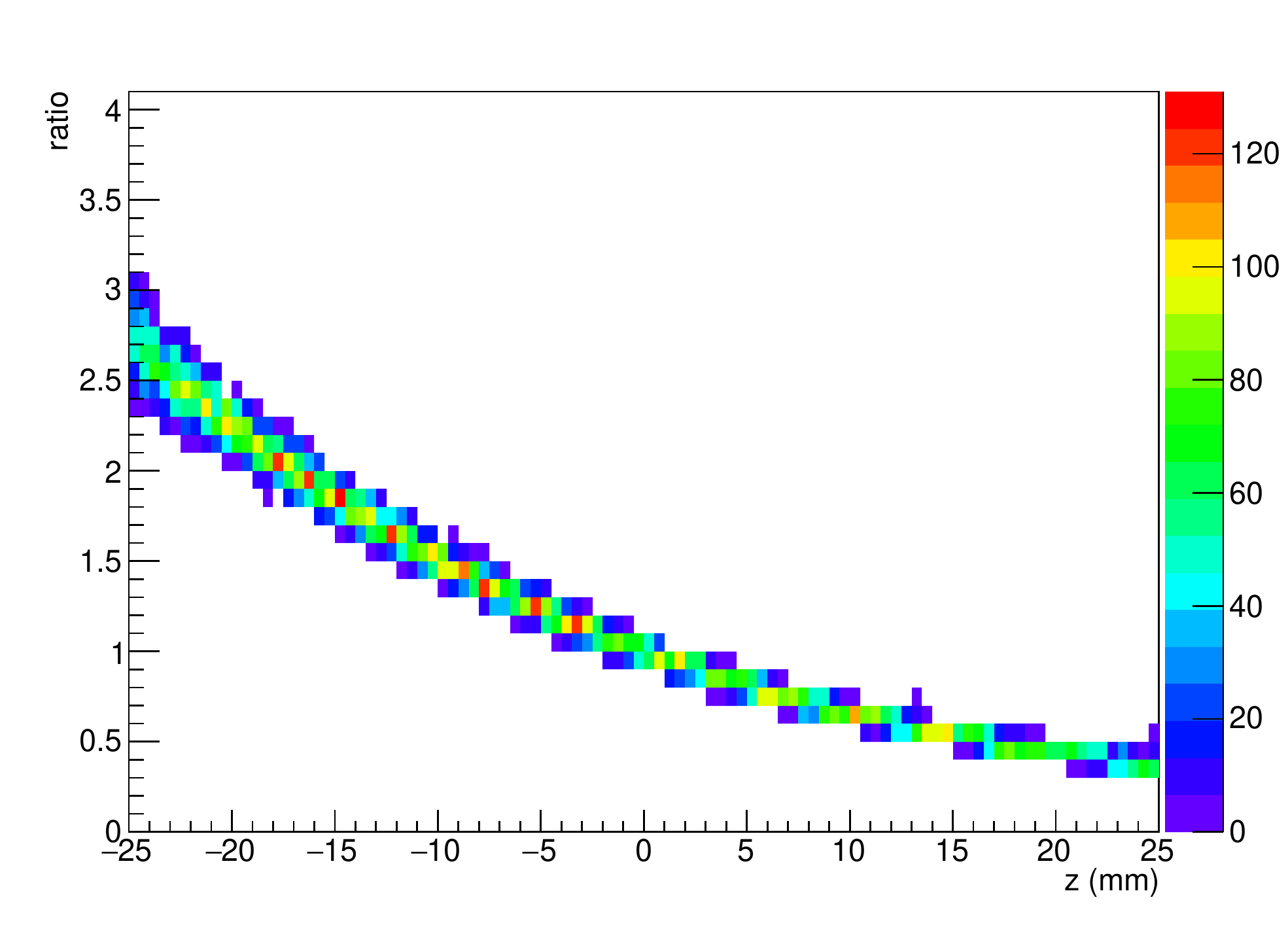}
	\caption{\label{fig.zratio}  Ratio of the signal in the entry and the exit face (the signal in a face is defined as the sum of the signals of all its SiPMs) as a function of the longitudinal coordinate.}
\end{figure}

Figure \ref{fig.zratio} shows the ratio of the signal in the entry and the exit face (the signal in a face is defined as the sum of the signals of all its SiPMs) as a function of the longitudinal coordinate. This ratio measures the longitudinal coordinate.  

%LXSC6_64

There is also some diffuse light that introduces noise in the detector, so if we want to compute barycenter properly, we need to apply some cuts. There are several approaches, we have tried two of them: (a) find the maximum SiPM in the plane and take only those SiPM having more than a percentage of its charge or (b) find the maximum SiPM and take those adjacent to it, building a cluster with the maximum on its center. The first approach has worked better for us, usually with a cut around 75\%.

\begin{table}[h]
\caption{\label{tab.position} Resolution (mm) in $x$ and $y$ coordinates for different LXSC configurations.}
\begin{center}
 \begin{tabular}{c|cc|cc|cc}
  \toprule
\multirow{2}{*}{Planes\textbackslash SiPM}  & \multicolumn{2}{c}{\textbf{36 SiPM}} & \multicolumn{2}{c}{\textbf{49 SiPM}} & \multicolumn{2}{c}{\textbf{64 SiPM}} \\
  \cline{2-7}
  & \textbf{x} & \textbf{y} & \textbf{x} & \textbf{y} & \textbf{x} & \textbf{y} \\
  \hline
    \textbf{2 Planes} & 2.1 & 2.1 & 1.8 & 1.8 & 1.9 & 1.9 \\
    \textbf{4 Planes} & 1.5 & 1.5 & 1.3 & 1.3 & 1.3 & 1.3 \\
    \textbf{6 Planes} & 0.9 & 0.9 & 0.8 & 0.7 & 0.8 & 0.8 \\
    \toprule
 \end{tabular}
\end{center}
\end{table}

\begin{table}[h]
\caption{\label{tab.positionZ} Resolution (mm) in $z$ coordinate for different LXSC configurations using the ratio entry/exit plane and using barycenter.}
\begin{center}
 \begin{tabular}{c|ccc}
  \toprule
  Planes\textbackslash SiPM & \textbf{36 SiPM} & \textbf{49 SiPM} & \textbf{64 SiPM} \\
   \hline
  \textbf{2 Planes (ratio)} & 1.5 & 1.3 & 1.2 \\
  \textbf{4 Planes (barycenter)} & 1.5 & 1.4 & 1.4 \\
  \textbf{6 Planes (barycenter)} & 0.9 & 0.8 & 0.9 \\
  \textbf{6 Planes (ratio)} & 1.2 & 1.0 & 1.0 \\
    \toprule
 \end{tabular}
\end{center}
\end{table}

\begin{figure}[!htb]
	\centering
	\subfloat[LXSC6\_64]{
		\includegraphics[width=.5\textwidth]{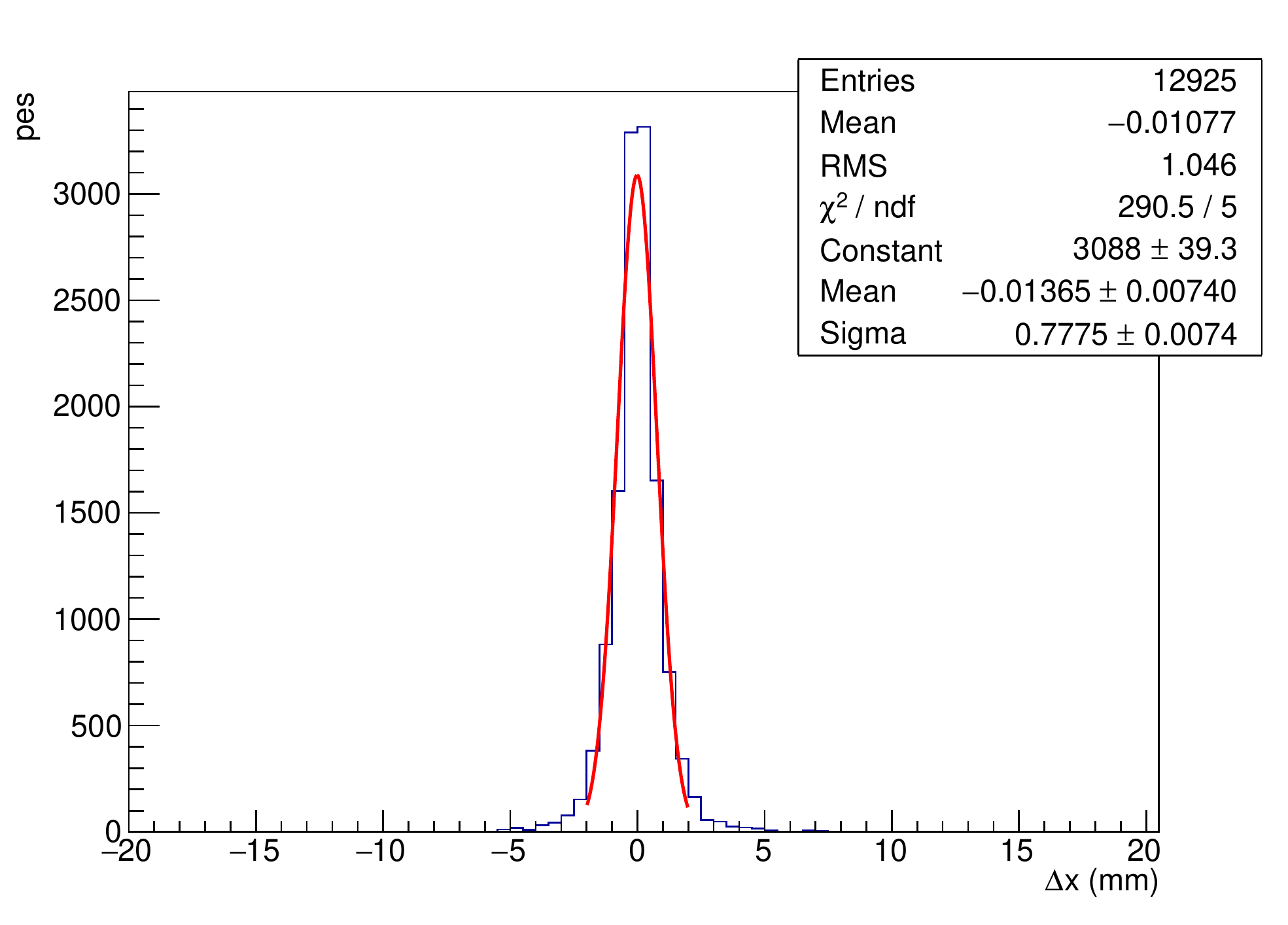}
	}
	\subfloat[LXSC2\_36]{
		\includegraphics[width=.5\textwidth]{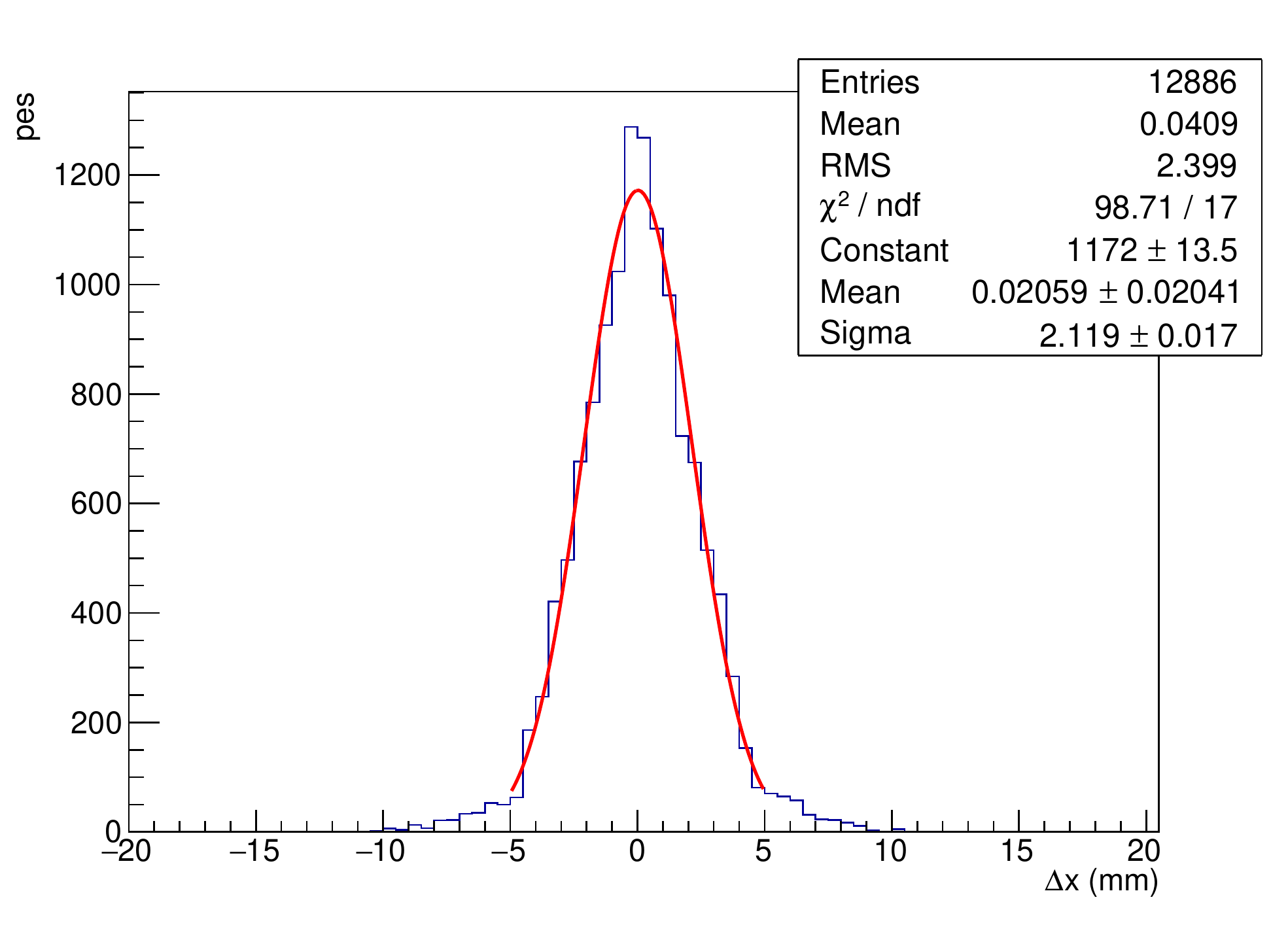}
	}
	\caption{\label{fig.xBest} Resolution in a transverse coordinate. (a) Best case, LXSC fully instrumented (6 planes with 64 SiPM each). (b) Worst case, most sparse configuration, 2 planes with 36 SiPM each.}
\end{figure}

%The resolution achieved with a box of $10\times10\times10$ cm$^3$ fully instrumented (LXSC10) is around 2.2 mm in the three coordinates. 
\newpage
Table \ref{tab.position} shows the results for both transverse coordinates for different $5\times5\times5$ cm$^3$ configurations. Figure \ref{fig.xBest} show the histograms for best (LXSC6-64) and worst cases (LXSC2-36).

In Table \ref{tab.positionZ} are shown the results for the longitudinal coordinate computed using barycenter or entry/exit ratio, it can be seen that the ratio works very well. 

As expected, reducing the amount of instrumentation leads to worse resolutions but the effect is quite mild, and appears as a good tradeoff for large PET scanners.  

Table \ref{tab.position2Z} shows how resolution changes if we reduce the longitudinal size of LXSC2. The transverse resolution improves by a factor of 2 for the thinner cell, while the longitudinal resolution worsens by a factor 40\%. However, the resolution in $z$~is still much better than the resolution that can be achieved by SSDs (the thickness of the detector over $\sqrt{12}$, typically $2/\sqrt{12} \sim 6 mm$~even for the thinner cell, and the improvement in transverse resolution may be relevant for small animal or brain PET. 

% It gets better but, as we have seen in previous sections, energy resolution is worse (see Table \ref{tab.energy2}) and, most important, many gammas cross the detector without interaction (see Table \ref{tab.lxscZ}). Therefore, we conclude that $5\times5\times5$ cm$^3$ is a good size for LXSC. The amount of instrumentation will ultimately be determined by the cost.
%
%We have also studied whether geometrical corrections are needed or not. Figure \ref{fig.bias} shows that near the edges (near the SiPMs) the reconstruction is better than in the middle, where the interaction point is far from both planes, but there is no bias we can correct geometrically.

\begin{table}[h]
\caption{\label{tab.position2Z} Resolution in the three coordinates for LXSC2 varying the longitudinal size.}
\begin{center}
 \begin{tabular}{c|ccc}
  \toprule
   {\bf Longitudinal size} & \textbf{$x$ coordinate} & \textbf{$y$ coordinate} & \textbf{$z$ coordinate}\\
   \hline
  {\bf 2 cm} & 1.0 & 1.0 & 1.8\\
  {\bf 3 cm} & 1.4 & 1.4 & 1.4\\
  {\bf 4 cm} & 1.7 & 1.7 & 1.4\\
  {\bf 5 cm} & 1.9 & 1.9 & 1.3\\
  \toprule
 \end{tabular}
\end{center}
\end{table}

%\begin{figure}[h]
%	\centering
%	\subfloat[$x$ coordinate]{
%		\includegraphics[width=.5\textwidth]{img/xbias.pdf}
%	}
%	\subfloat[$y$ coordinate]{
%		\includegraphics[width=.5\textwidth]{img/ybias.pdf}
%	}
%	\caption{\label{fig.bias} Difference between reconstructed position and true position for both transverse coordinates as a function of the longitudinal one.}
%\end{figure}

Summarizing, for the standard box dimensions, the space resolution obtained for the LXSC6 is better than 1 mm r.m.s. The resolution in the LXSC2 is better than 2 mm r.m.s. in the transverse coordinates ($x,y$) and 1.5 mm in the longitudinal coordinate. The transverse resolution can be improved by a substantial factor of 2 by reducing the thickness of the box to 2 cm. 

\subsection{Neural network}
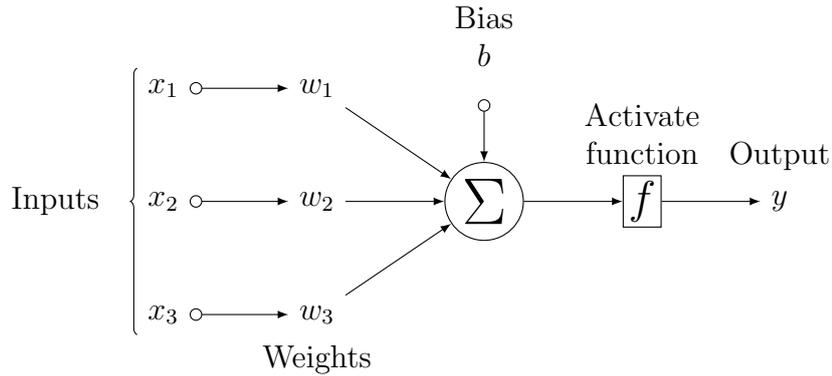
\begin{figure}[!h]
	\centering
\begin{tikzpicture}[
init/.style={
  draw,
  circle,
  inner sep=2pt,
  font=\Huge,
  join = by -latex
},
squa/.style={
  draw,
  inner sep=2pt,
  font=\Large,
  join = by -latex
},
start chain=2,node distance=13mm
]
\node[on chain=2] 
  (x2) {$x_2$};
\node[on chain=2,join=by o-latex] 
  {$w_2$};
\node[on chain=2,init] (sigma) 
  {$\displaystyle\Sigma$};
\node[on chain=2,squa,label=above:{\parbox{2cm}{\centering Activate \\ function}}]   
  {$f$};
\node[on chain=2,label=above:Output,join=by -latex] 
  {$y$};
\begin{scope}[start chain=1]
\node[on chain=1] at (0,1.5cm) 
  (x1) {$x_1$};
\node[on chain=1,join=by o-latex] 
  (w1) {$w_1$};
\end{scope}
\begin{scope}[start chain=3]
\node[on chain=3] at (0,-1.5cm) 
  (x3) {$x_3$};
\node[on chain=3,label=below:Weights,join=by o-latex] 
  (w3) {$w_3$};
\end{scope}
\node[label=above:\parbox{2cm}{\centering Bias \\ $b$}] at (sigma|-w1) (b) {};

\draw[-latex] (w1) -- (sigma);
\draw[-latex] (w3) -- (sigma);
\draw[o-latex] (b) -- (sigma);

\draw[decorate,decoration={brace,mirror}] (x1.north west) -- node[left=10pt] {Inputs} (x3.south west);
\end{tikzpicture}
	\caption{\label{fig.nnNode} Node of neural network.}
\end{figure}

Machine learning can also be used to reconstruct event positions. We have explored this possibility using neural networks, which is a set of connected nodes called {\it neurons}. Each one of them receives some numerical inputs, $x_i$, that have a weight associated, $w_i$. With this data the activation function is computed giving the output for one node: $y = f(\sum_i w_i x_i )$. We have used a common activation function, the logistic function: $f(x) = e^x/(1+e^x)$. The idea is illustrated on Figure \ref{fig.nnNode}. 

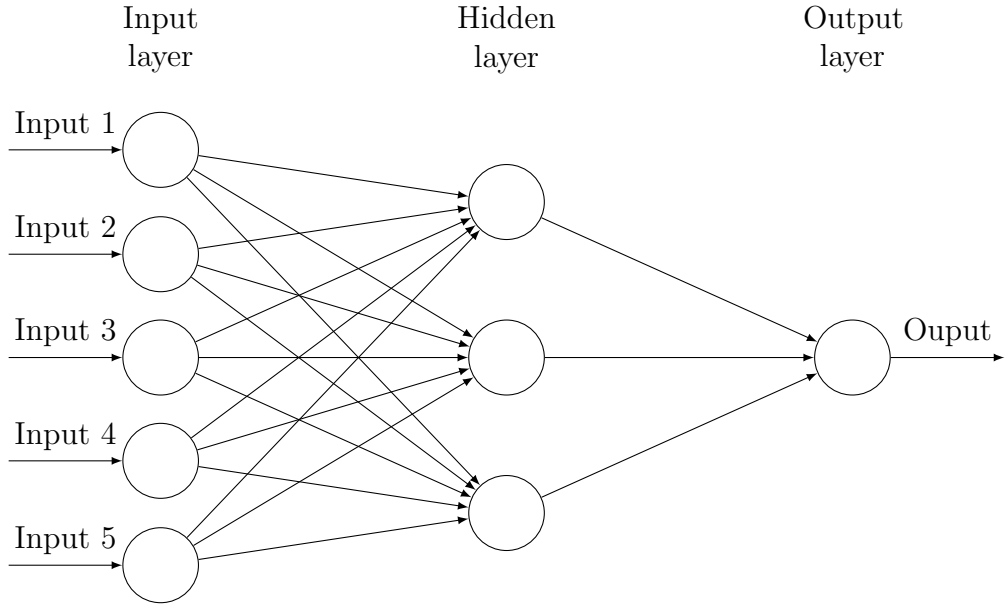
\begin{figure}[!htb]
	\centering
	\begin{tikzpicture}[
	plain/.style={
	  draw=none,
	  fill=none,
	  },
	net/.style={
	  matrix of nodes,
	  nodes={
	    draw,
	    circle,
	    inner sep=10pt
	    },
	  nodes in empty cells,
	  column sep=2cm,
	  row sep=-9pt
	  },
	>=latex
	]
	\matrix[net] (mat)
	{
	|[plain]| \parbox{1.3cm}{\centering Input\\layer} & |[plain]| \parbox{1.3cm}{\centering Hidden\\layer} & |[plain]| \parbox{1.3cm}{\centering Output\\layer} \\
	& |[plain]| \\
	|[plain]| & \\
	& |[plain]| \\
	|[plain]| & |[plain]| \\
	& & \\
	|[plain]| & |[plain]| \\
	& |[plain]| \\
	|[plain]| & \\
	& |[plain]| \\
	};
	\foreach \ai [count=\mi ]in {2,4,...,10}
	  \draw[<-] (mat-\ai-1) -- node[above] {Input \mi} +(-2cm,0);
	\foreach \ai in {2,4,...,10}
	{\foreach \aii in {3,6,9}
	  \draw[->] (mat-\ai-1) -- (mat-\aii-2);
	}
	\foreach \ai in {3,6,9}
	  \draw[->] (mat-\ai-2) -- (mat-6-3);
	\draw[->] (mat-6-3) -- node[above] {Ouput} +(2cm,0);
	\end{tikzpicture}
	\caption{\label{fig.nn} Example Neural Network. All inputs are connected to all nodes in next layer (the hidden layer) and the hidden layer is connected to the output. We have use only only one hidden layer and one output.}
\end{figure}

The basic architecture of a feed-forward neural network is illustrated in Figure \ref{fig.nn}. There is one input layer with as many nodes as inputs there are to the problem, one or more hidden layers, each one with a number of hidden nodes and, finally, at the end, the output(s). All nodes in one layer are connected to those of the next layer. The input nodes only pass the values $x_i$ to the first hidden layer, each hidden node will compute its activation function $f_h$ and then pass the result to next level until the output(s) node(s), which will compute the final answer. Each connection between nodes will have a weight, which initially will be a random value. Therefore the output for a network with only one hidden layer will be:

$$ y = f_o \left(b_k + \sum_h w_{hk} f_h \left(b_h + \sum_i w_{ih}x_i\right) \right)$$
where $f_o$ is the output activation function, $w_{ij}$ are the weights from layer $i$ to layer $j$ and $b_j$ are {\it biases}, numbers that are added by each node to improve learning process.

This kind of techniques requires some {\it training}, we need a set of inputs for which we know the correct output to adjust the parameters of the network. The way this is done is using the backpropagation algorithm, which computes the output of the network
and make small corrections to the weights to minimize the error. This is repeated for all inputs in the training set, one iteration over all inputs is called an {\it epoch}. The algorithm stops when the error is less than a threshold  or when a maximum number of epochs is achieved. To measure the error the easiest way is using least-squares method.

We have tested this approach to reconstruct both transverse coordinates ($x$ and $y$) using as inputs all the SiPM values from the entry plane of a LXSC2 simulation. To do this we have trained two networks, one for each coordinate. The architecture chosen has 64 inputs nodes, one hidden layer with 100 nodes and one output. We have a dataset with 12925 photoelectric events and we divided it between 6463 for training set and 6462 for test set. After training the network we have compute the resolution for both sets. 

\begin{figure}[!htb]
	\centering
	\subfloat[Train set]{
		\includegraphics[width=.5\textwidth]{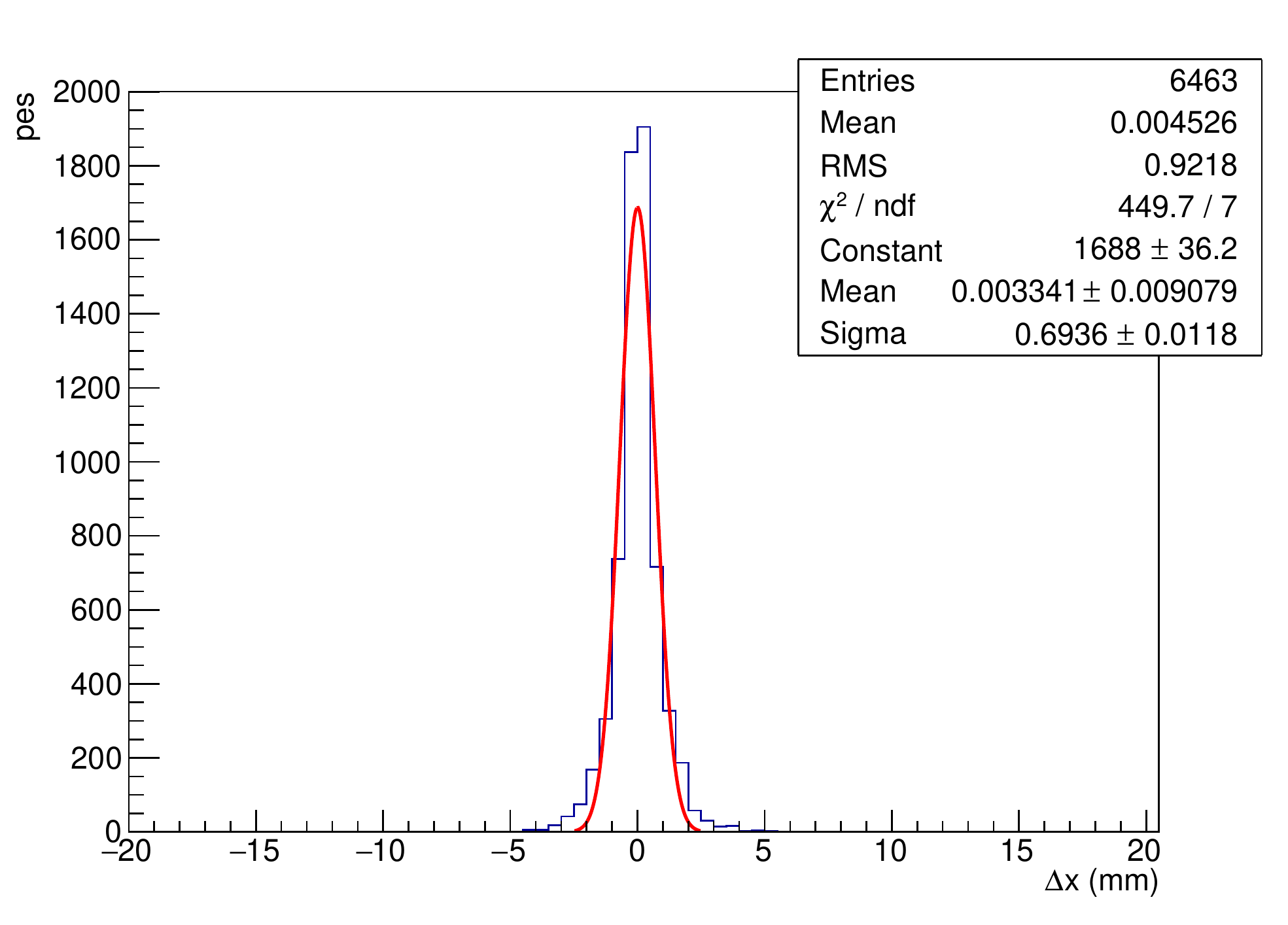}
		\label{fig.nnPlotTrain}
	}
	\subfloat[Test set]{
		\includegraphics[width=.5\textwidth]{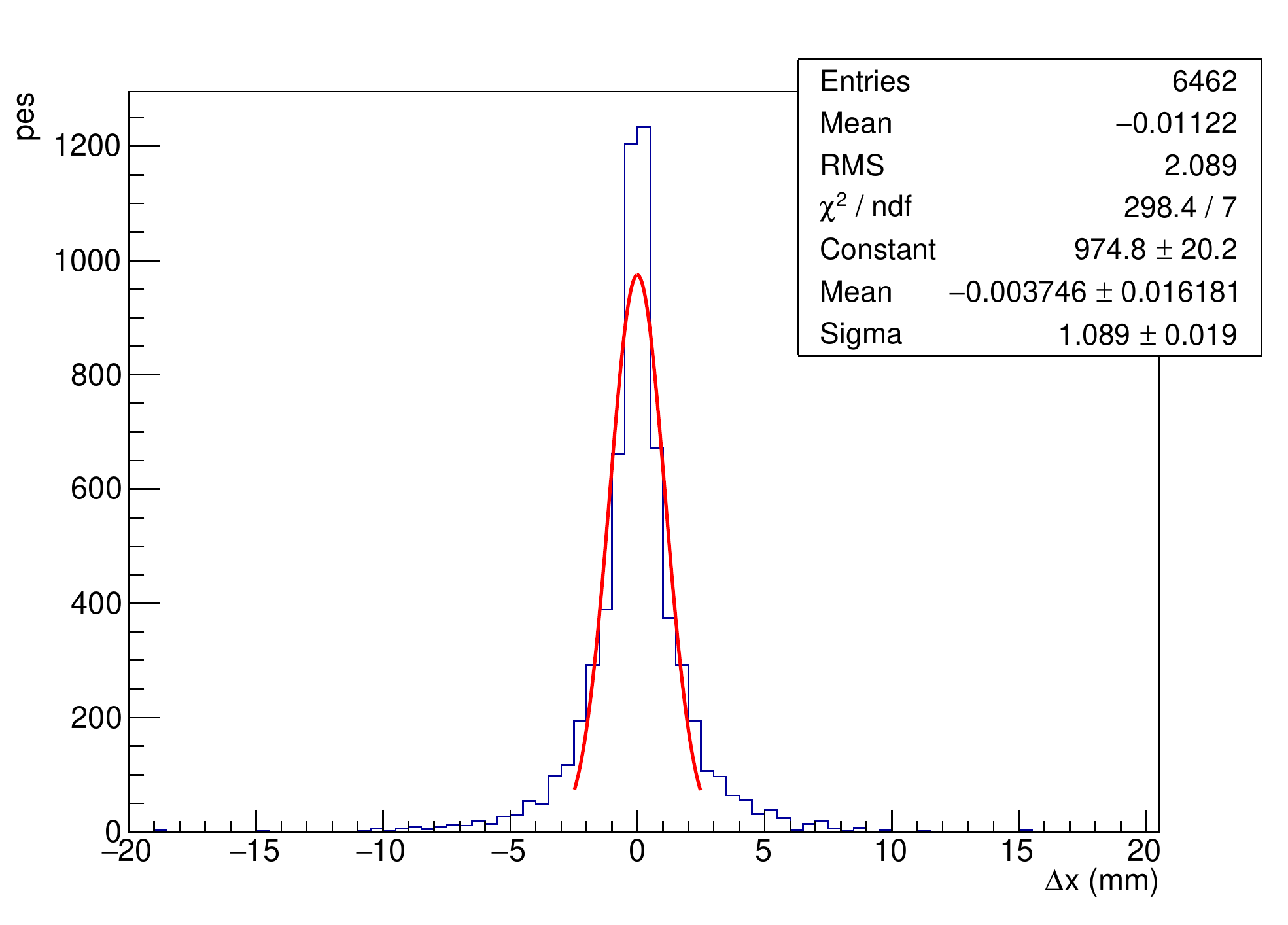}
		\label{fig.nnPlotTest}
	}
	\caption{\label{fig.nnPlot} Resolution in $x$ coordinate using a neural network to reconstruct position. (a) Resolution for train set. (b) Resolution for test set.}
\end{figure}

Results seem very promising as is shown by Figure \ref{fig.nnPlot}. There seems to be a little overfitting, but a good result overall. The histogram for test set is not completely gaussian but the majority of events has been well reconstructed. For $y$ coordinate we get similar results, having obtained resolutions of 0.652 and 1.051 mm for train and test respectively. Clearly, neural networks, and machine learning in a broader sense, seems like a very promising direction to further explore in this problem.

\section{PETALO Scanners}
\label{sec.pets}

\begin{figure}[!htb]
	\centering
	\includegraphics[scale=0.4]{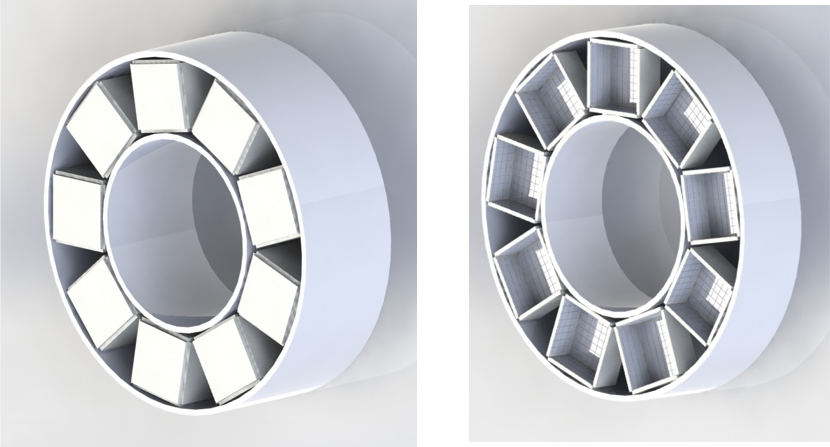}
	\caption{\label{fig.smallPet} A conceptual drawing of a small brain PET, based in 9 LXSC. }
\end{figure}

As we have shown in previous sections, the excellent energy resolution, excellent CRT and good spatial resolution of the LXSC makes it the cornerstone of a number of PET systems.  For example:

\begin{enumerate}
\item {\bf A ``full body'' PET} application (e.g, a large PET device, comprising several rings of large diameter, covering the full torso of the patient), needs to be optimized for cost, and could benefit for the most sparse configuration (LXSC2 of $5\times5\times5$~cm$^3$~ with 36 SiPMs of 6 mm$^2$~per face).
\item {\bf A ``small animal'' PET} application, which intends to reconstruct images of the small organs of test animals such as mice, needs to be optimized for spatial resolution, and could be based in a LXSC2 fully instrumented with small SiPMs (of 3 mm$^2$~or smaller). The size of the box should be adjusted to the PET smallish diameter (to keep good packing) and the thickness of the box could be reduced to 2 cm, to optimize spatial resolution (trading it for efficiency which could be less critical for this application).
\item {\bf A ``brain scan'' PET} application, requires less modules than a full body PET and does not require a resolution as good as small animal PET, thus it could be based in an LXSC2 of a size intermediate between the small animal PET and the full body PET.
\end{enumerate}

%
% It has better energy and spatial resolutions than PETs currently in the market and it is also cheaper. Figure \ref{fig.smallPet} shows the design of a small brain PET using LXSC.
%
%In this section we discuss another two important features that PETALO potentially has: TOF-PET capability and NMR compatibility.
\subsubsection*{TOF application}

The quick decay time of xenon and the fast circuitry available in modern SiPMs offers an extraordinary potential for TOF applications. About 2.5\% of the photons are emitted in LXe in the first 50 ps after the interaction. The SiPM recording more signal in one event sees typically  2\% of the total photoelectrons (pes), and therefore it records 5 pes in the first 50 ps, enough to trigger a signal. For comparison, one nanosecond is needed in LSO to emit 2.5\% of the photons. Commercial TOF-PET systems based in LYSO (a proprietary version of LSO) have achieved a system time resolution of 600 ps. Since LXe features {\em both} higher light yield and much faster time response, a system resolution at the level or better than 200 ps appears possible. Thus, PETALO could represent a breakthrough in the field of PET-TOF. 

\subsubsection*{MRI compatibility}

Magnetic Resonante Imaging (MRI) is a medical imaging technique that shows anatomical features using an NMR apparatus. There is strong interest in building a MRI-compatible PET scanner capable of acquiring PET images simultaneously with MRI images and, therefore, combining anatomical and functional information.

The main issue is that an MRI requires strong magnetic fields and, hence, is not compatible with current PET systems due to the use of photomultiplier tubes (PMT), which will not function under high magnetic fields.

On the other hand, the LXSC is built using non-magnetic materials, and unlike PMTs, SiPMs can operate in very high magnetic fields. Thus, PETALO can operate inside the very intense magnetic field generated by NMR devices. Furthermore, a NMR apparatus requires a large cryostat which can also accommodate the LXSC modules that make up PETALO. The technology offers, therefore, the possibility of building a fully MRI compatible device.

%
%\newpage
%\thispagestyle{newstyle}

%\newpage\null\thispagestyle{empty}\newpage
%\thispagestyle{newstyle}
\newpage
\section{Conclusions}
\label{sec.conclu}

PETALO is a new technology for TOF-PET systems based on the LXSC, a fully hermetic, homogenous box filled with liquid xenon and equipped with SiPM coated with TPB. PETALO offers the following advantages:

\begin{enumerate}
\item Light yield higher than any conventional SSD.
\item Excellent intrinsic energy resolution (3.5 -- 4.5 \% FWHM depending on configuration of LXSC and of light yield). 
\item Excellent spatial resolution (1-2 mm in the three coordinates).
\item Potentially capable of detecting multi-site Compton events. Thus suitable as Compton telescope.
\item Very fast time response, resulting in enhanced sensitivity (reduced number of random coincidences) and making it possible breakthrough TOF application. 
\item Fully MRI compatible. 
\item Competitive cost. The sparse version of the LXSC2 would cost today roughly 2-3 times less than the cost of the equivalent LSO unit. With the cost of SiPMs falling, the  cost of PET scanners will soon be fully dominated by the material of choice. Xenon is much cheaper than LSO and thus a large-scale apparatus (full body PET) is conceivable. 
\end{enumerate}

With respect to the pioneer work of the Waseda group, PETALO introduces the concept of the fully reflective, hermetic and homogenous LXSC, capable to detect the VUV light emitted by xenon with high efficiency and very small geometrical effects, thanks to the use of SiPMs. 

%The problem of identifying Compton events and reconstructing more than one vertex remains open. Solving it would give us all the potential of the LXSC and would also be a big improvement in PET technologies.

\newpage\null\thispagestyle{empty}

\end{document}